\RequirePackage{fix-cm}
\documentclass[smallcondensed]{svjour3}     
\smartqed  
\usepackage[numbers]{natbib}
\usepackage{booktabs}
\usepackage{url}
\usepackage{amssymb}
\usepackage{amsmath}
\usepackage{wrapfig}
\usepackage{enumitem}
\usepackage{mdwlist}
\usepackage{setspace}
\usepackage[caption=false,font=footnotesize]{subfig}
\usepackage{microtype}
\usepackage{color}
\usepackage{algorithm,algorithmic}
\usepackage{tikz}
\usepackage{rotating}
\usepackage{makecell}
\usepackage{multirow}
\usepackage{dsfont}
\usetikzlibrary{arrows,patterns,automata,backgrounds,shadows,decorations,decorations,fit,petri,positioning,petri,shapes,calc,graphs,arrows.meta,shadows}
\tikzset{place/.append style={circle,draw=black,thick,inner sep=0pt,minimum size=3mm,label position=below}}
\tikzset{transition/.append style={rectangle,draw=black,thick,inner sep=1pt,minimum size=4mm}}
\tikzset{every edge/.append style={-{>[sep=0pt]}, thin}}
\tikzset{pre/.append style={<-,shorten <=0pt,shorten >=0pt}}
\tikzset{post/.append style={->,shorten >=0pt,shorten <=0pt}}

\newlength{\hatchspread}
\newlength{\hatchthickness}
\newlength{\hatchshift}
\newcommand{\hatchcolor}{}
\tikzset{hatchspread/.code={\setlength{\hatchspread}{#1}},
	hatchthickness/.code={\setlength{\hatchthickness}{#1}},
	hatchshift/.code={\setlength{\hatchshift}{#1}},
	hatchcolor/.code={\renewcommand{\hatchcolor}{#1}}}
\tikzset{hatchspread=3pt,
	hatchthickness=0.4pt,
	hatchshift=0pt,
	hatchcolor=black}
\pgfdeclarepatternformonly[\hatchspread,\hatchthickness,\hatchshift,\hatchcolor]
{custom north west lines}
{\pgfqpoint{\dimexpr-2\hatchthickness}{\dimexpr-2\hatchthickness}}
{\pgfqpoint{\dimexpr\hatchspread+2\hatchthickness}{\dimexpr\hatchspread+2\hatchthickness}}
{\pgfqpoint{\dimexpr\hatchspread}{\dimexpr\hatchspread}}
{
	\pgfsetlinewidth{\hatchthickness}
	\pgfpathmoveto{\pgfqpoint{0pt}{\dimexpr\hatchspread+\hatchshift}}
	\pgfpathlineto{\pgfqpoint{\dimexpr\hatchspread+0.15pt+\hatchshift}{-0.15pt}}
	\ifdim \hatchshift > 0pt
	\pgfpathmoveto{\pgfqpoint{0pt}{\hatchshift}}
	\pgfpathlineto{\pgfqpoint{\dimexpr0.15pt+\hatchshift}{-0.15pt}}
	\fi
	\pgfsetstrokecolor{\hatchcolor}
	\pgfusepath{stroke}
}

\newcommand{\fleche}{\longrightarrow}
\newcommand{\flsup}[1]{\stackrel{#1}{\fleche}}
\newcommand{\step}[1]{\flsup{#1}}           
\newcommand{\Lan}                 {\mathfrak{L}}

\newcommand{\pre}[1]{\bullet #1}
\newcommand{\APN}{\mathit{APN}}

\newcommand{\MF}{\mathit{MF}}

\newcolumntype{C}[1]{>{\centering\let\newline\\\arraybackslash\hspace{0pt}}m{#1}}

\begin{document}

\title{Discovering More Precise Process Models from Event Logs by Filtering Out Chaotic Activities}


\author{Niek Tax \and Natalia Sidorova \and Wil M. P. van der Aalst
}


\institute{
	Department of Mathematics and Computer Science\\
	Eindhoven University of Technology\\
	P.O. Box 513, 5600MB Eindhoven, The Netherlands\\
	Email: \email{\{n.tax,n.sidorova,w.m.p.v.d.aalst\}@tue.nl}
}

\date{Received: date / Accepted: date}

\maketitle

\begin{abstract}
\emph{Process Discovery} is concerned with the automatic generation of a process model that describes a business process from execution data of that business process. Real life event logs can contain \emph{chaotic activities}. These activities are independent of the state of the process and can, therefore, happen at rather arbitrary points in time. We show that the presence of such \emph{chaotic activities} in an event log heavily impacts the quality of the process models that can be discovered with process discovery techniques. The current modus operandi for filtering activities from event logs is to simply filter out infrequent activities. We show that frequency-based filtering of activities does not solve the problems that are caused by chaotic activities. Moreover, we propose a novel technique to filter out chaotic activities from event logs. We evaluate this technique on a collection of seventeen real-life event logs that originate from both the business process management domain and the smart home environment domain. As demonstrated, the developed activity filtering methods enable the discovery of process models that are more behaviorally specific compared to process models that are discovered using standard frequency-based filtering.
\end{abstract}

\section{Introduction}
\begin{figure}[b]
	\centering
	\hspace{-0.3cm}
	\scalebox{0.7}{
		\subfloat[\label{sfig:traces_example}]{
			\begin{tabular}{l}
				\toprule
				Event sequences \\
				\midrule
				$\langle$A,B,D,E,H$\rangle$\\
				$\langle$A,D,C,E,G$\rangle$\\
				$\langle$A,C,D,E,F,B,D,E,G$\rangle$\\
				$\langle$A,D,B,E,H$\rangle$\\
				$\langle$A,C,D,E,F,D,C,E,F,C,D,E,H$\rangle$\\
				$\langle$A,C,D,E,G$\rangle$\\
				\bottomrule
			\end{tabular}
		}
		\hspace{-0.35cm}
		\subfloat[\label{sfig:local_model_example}]{
			\scalebox{0.95}{
				\raisebox{-.5\height}{
					\begin{tikzpicture}
					[node distance=0.7cm,
					on grid,>=stealth',
					bend angle=20,
					auto,
					every place/.style= {minimum size=4mm},
					every transition/.style = {minimum size = 3mm}
					]
					\node [place, tokens = 1] at (-1.4,0) (p0){};
					\node [transition] (ts4) [minimum width=3mm] at (-0.7,0) {A}
					edge[pre] node[auto] {} (p0);
					\node [place] at (0,0) (p1){}
					edge[pre] node[auto] {} (ts4);
					\node [place] at (1.4,0.35) (p2){};
					\node [place] at (1.4,-0.35) (p22){};
					\node [place] at (2.8,0.35) (p3) {};
					\node [place] at (2.8,-0.35) (p33) {};
					
					\node [transition] (t1) [minimum width=3mm,fill=lightgray] at (0.7,0.0){}
					edge[pre] node[auto] {} (p1)
					edge[post] node[auto] {} (p2)
					edge[post] node[auto] {} (p22);
					
					\node [transition] (t2) [align=center] at (2.1,0.7) {B}
					edge [pre] node[auto] {} (p2)
					edge[post] node[auto] {} (p3);
					\node [transition] (t3) [] at (2.1,0){C}
					edge[pre] node[auto] {} (p2)
					edge[post] node[auto] {} (p3);
					
					\node [transition] (t3) [] at (2.1,-.5){D}
					edge[pre] node[auto] {} (p22)
					edge[post] node[auto] {} (p33);
					
					\node [transition] (ts5) [minimum width=3mm] at (3.5,0.0) {E}
					edge[pre] node[auto] {} (p33)
					edge[pre] node[auto] {} (p3);
					\node [place] at (4.2,0) (p4) {}
					edge[pre] node[auto] {} (ts5);
					
					\node [transition] (t6) [minimum width=3mm] at (4.9,0.35) {G}
					edge[pre] node[auto] {} (p4);
					\node [transition] (t7) [minimum width=3mm] at (4.9,-0.35) {H}
					edge[pre] node[auto] {} (p4);
					\node [place,pattern=custom north west lines,hatchspread=1.5pt,hatchthickness=0.25pt,hatchcolor=gray] at (5.6,0) (p5) {}
					edge[pre] node[auto] {} (t6)
					edge[pre] node[auto] {} (t7);
					
					\node [transition] (ts3) [minimum width=3mm] at (4.2,-0.7) {F}
					edge[pre] node[auto] {} (p4);
					\draw [post] (ts3) to [out=180,in=0] ($(t3)-(0,0.4)$) to [out=180,in=-90 ] (p1);
					\end{tikzpicture}}}
		}
	}
	\caption{\emph{(a)} Event log with A=register request, B=examine thoroughly, C=examine casually, D=check ticket, E=decide, F=re-initiate request, G=pay compensation, H=reject request, and \emph{(b)} the Petri net mined from this log with the Inductive Miner \cite{Leemans2013}.}
	\label{fig:unstructured_log_discovery}
	\vspace{-0.1cm}
\end{figure}
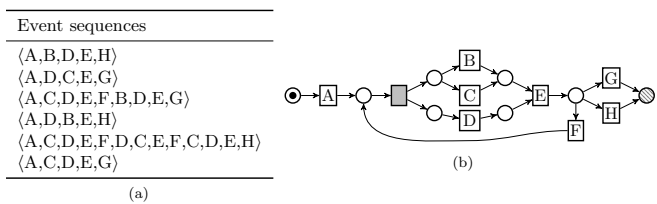
\emph{Process Mining} \cite{Aalst2016} is a scientific discipline that bridges the gap between process analytics and data analysis and focuses on the analysis of event data logged during the execution of a business process. Events contain information on what was done, by whom, for whom, where, when, etc. Such event data is often readily available from information systems such as ERP, CRM, or BPM systems. \emph{Process discovery}, which plays a prominent role in process mining, is the task of automatically generating a process model that accurately describes a business process based on such event data. Many process discovery techniques have been developed over the last decade (e.g. \cite{Buijs2012,Goedertier2009,Gunther2007,Herbst2000,Leemans2013,Sole2013,Zelst2015}), producing process models in various forms, such as Petri nets~\cite{Murata1989}, process trees \cite{Buijs2012}, and BPMN models~\cite{OMG2011}.

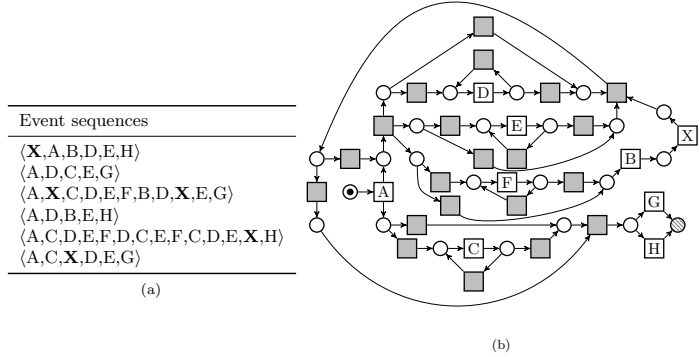
\begin{figure}[t]
	\centering
	\hspace{-0.3cm}
	\scalebox{0.7}{
		\subfloat[\label{sfig:traces_example_noisy}]{
			\begin{tabular}{l}
				\toprule
				Event sequences \\
				\midrule
				$\langle$\textbf{X},A,B,D,E,H$\rangle$\\
				$\langle$A,D,C,E,G$\rangle$\\
				$\langle$A,\textbf{X},C,D,E,F,B,D,\textbf{X},E,G$\rangle$\\
				$\langle$A,D,B,E,H$\rangle$\\
				$\langle$A,C,D,E,F,D,C,E,F,C,D,E,\textbf{X},H$\rangle$\\
				$\langle$A,C,\textbf{X},D,E,G$\rangle$\\
				\bottomrule
			\end{tabular}
		}
		\hspace{-0.35cm}
		\raisebox{-2.6cm}{
			\scalebox{0.895}{
				\subfloat[\label{sfig:local_model_example_noisy}]{
					\begin{tikzpicture}%
					[node distance=0.7cm,
					on grid,>=stealth',
					bend angle=20,
					auto,
					every place/.style= {minimum size=4mm},
					every transition/.style = {minimum size = 3mm}
					]
					\node [place,tokens=1](source 1) [] {}; 
					
					\node [transition](A) [right of=source 1] {A};
					\node [place](source 5) [below of=A] {}; 
					\node [place](source 9) [above of=A] {}; 
					\node [transition](tau split) [above of=source 9,fill=lightgray] {}; 
					\node [place](source 12) [above of=tau split] {}; 
					\node [transition](tau start2) [right of=source 12, fill=lightgray] {}; 
					\node [place](replacement source 15) [right of=tau start2] {}; 
					\node [transition](D) [right of=replacement source 15] {D};
					\node [place](middle 14) [right of=D] {}; 
					\node [transition](tau from tree6) [right of=middle 14,fill=lightgray] {}; 
					\node [transition](tau from tree5) [above of=D,fill=lightgray] {}; 
					\node [place](sink 13) [right of=tau from tree6] {}; 
					\node [transition](tau from tree4) [above of=tau from tree5,fill=lightgray] {}; 
					
					\node [place](source 16) [right of=tau split] {}; 
					\node [transition](tau start3) [right of=source 16,fill=lightgray] {}; 
					\node [place](replacement source 19) [right of=tau start3] {}; 
					\node [transition](E) [right of=replacement source 19] {E};
					\node [place](middle 18) [right of=E] {}; 
					\node [transition](tau from tree9) [right of=middle 18,fill=lightgray] {}; 
					\node [place](sink 17) [right of=tau from tree9] {}; 
					\node [transition](tau from tree8) [below of=E,fill=lightgray] {}; 
					\node [transition](tau from tree7) [left of=tau from tree8,fill=lightgray] {}; 
					
					\node [place](source 20) [below of=source 16] {}; 
					\node [transition](tau start4) [below right of=source 20,fill=lightgray] {}; 
					\node [place](replacement source 24) [right of=tau start4] {}; 
					\node [transition](F) [right of=replacement source 24] {F};
					\node [place](middle 23) [right of=F] {}; 
					\node [transition](tau from tree12) [right of=middle 23,fill=lightgray] {}; 
					\node [place](sink 22) [right of=tau from tree12] {}; 
					\node [transition](B) [above right of=sink 22] {B};
					\node [place](sink 25) [right of=B] {}; 
					\node [transition](X) [above right of=sink 25] {X};
					\node [place](sink 21) [above left of=X] {}; 
					\node [transition](tau from tree11) [below left of=middle 23,fill=lightgray] {}; 
					
					\node [transition](tau join2) [right of=sink 13,fill=lightgray] {}; 
					
					\node [place](middle 11) [left = 1.4cm of source 9] {}; 
					\node [transition](tau from tree13) [right of=middle 11,fill=lightgray] {}; 
					\node [transition](tau from tree10) [below left of=replacement source 24,fill=lightgray] {}; 
					\node [transition](tau from tree14) [below of=middle 11,fill=lightgray] {}; 
					\node [place](sink 10) [below of=tau from tree14] {}; 
					
					\node [transition](tau start) [below right of=source 5, fill=lightgray] {}; 
					\node [place](replacement source 8) [right of=tau start] {}; 
					\node [transition](C) [right of=replacement source 8] {C};
					\node [place](middle 7) [right of=C] {}; 
					\node [transition](tau from tree2) [below of=C, fill=lightgray] {}; 
					\node [transition](tau from tree3) [right of=middle 7,fill=lightgray] {}; 
					\node [place](sink 6) [above right of=tau from tree3] {}; 
					\node [transition](tau join) [right of=sink 6,fill=lightgray] {}; 
					\node [place](sink 4) [right of=tau join] {}; 
					\node [transition](G) [above right of=sink 4] {G};
					\node [transition](H) [below right of=sink 4] {H};
					\node [place,pattern=custom north west lines,hatchspread=1.5pt,hatchthickness=0.25pt,hatchcolor=gray](sink 2) [above right of=H] {}; 
					\node [transition](tau from tree) [right of=source 5,fill=lightgray] {}; 
					\draw
					(middle 18) edge[->] (tau from tree8)
					(D) edge[->] (middle 14)
					(tau start3) edge[->] (replacement source 19)
					(middle 14) edge[->] (tau from tree6)
					(tau from tree3) edge[->] (sink 6)
					(middle 7) edge[->] (tau from tree2)
					(sink 10) edge[->,bend right=55,distance=2.5cm] (tau join)
					(tau from tree9) edge[->] (sink 17)
					(A) edge[->] (source 5)
					(tau join2) edge[->,bend right=55,distance=4cm] (middle 11)
					(tau split) edge[->] (source 12)
					(tau from tree4) edge[->] (sink 13)
					(source 16) edge[->] (tau start3)
					(tau from tree12) edge[->] (sink 22)
					(A) edge[->] (source 9)
					(tau from tree13) edge[->] (source 9)
					(middle 23) edge[->] (tau from tree11)
					(sink 22) edge[->] (B)
					(tau split) edge[->] (source 20)
					(tau start) edge[->] (replacement source 8)
					(middle 23) edge[->] (tau from tree12)
					(middle 18) edge[->] (tau from tree9)
					(tau from tree7) edge[->,bend right=52,distance=0.51cm] (sink 17)
					(H) edge[->] (sink 2)
					(tau from tree8) edge[->] (replacement source 19)
					(source 20) edge[->,bend right=32,distance=0.51cm] (tau from tree10)
					(sink 4) edge[->] (H)
					(replacement source 19) edge[->] (E)
					(sink 21) edge[->] (tau join2)
					(tau from tree10) edge[->,bend right=52,distance=0.51cm] (sink 22)
					(source 12) edge[->] (tau start2)
					(E) edge[->] (middle 18)
					(source 12) edge[->] (tau from tree4)
					(sink 6) edge[->] (tau join)
					(tau start2) edge[->] (replacement source 15)
					(sink 25) edge[->] (X)
					(source 9) edge[->] (tau split)
					(G) edge[->] (sink 2)
					(tau from tree14) edge[->] (sink 10)
					(F) edge[->] (middle 23)
					(middle 14) edge[->] (tau from tree5)
					(source 5) edge[->] (tau start)
					(tau start4) edge[->] (replacement source 24)
					(source 16) edge[->] (tau from tree7)
					(replacement source 24) edge[->] (F)
					(source 5) edge[->] (tau from tree)
					(tau from tree5) edge[->] (replacement source 15)
					(source 1) edge[->] (A)
					(sink 4) edge[->] (G)
					(sink 17) edge[->] (tau join2)
					(X) edge[->] (sink 21)
					(sink 13) edge[->] (tau join2)
					(source 20) edge[->] (tau start4)
					(replacement source 15) edge[->] (D)
					(tau from tree2) edge[->] (replacement source 8)
					(tau split) edge[->] (source 16)
					(C) edge[->] (middle 7)
					(tau join) edge[->] (sink 4)
					(middle 7) edge[->] (tau from tree3)
					(tau from tree6) edge[->] (sink 13)
					(tau from tree) edge[->] (sink 6)
					(replacement source 8) edge[->] (C)
					(middle 11) edge[->] (tau from tree14)
					(B) edge[->] (sink 25)
					(tau from tree11) edge[->] (replacement source 24)
					(middle 11) edge[->] (tau from tree13);
					\end{tikzpicture}
				}			
	}}}
	\caption{\emph{(a)} The event log from Figure \ref{sfig:traces_example} with an added chaotic activity X, and \emph{(b)} the Petri net mined from this log with the Inductive Miner \cite{Leemans2013}.\looseness=-1}
	\label{fig:unstructured_log_discovery2}
	\vspace{-0.1cm}
\end{figure}

Figure \ref{sfig:local_model_example} shows an example process model from~\cite{Aalst2016} that describes a compensation request process. The process model consists of eight process steps (called activities): \emph{(A) register request}, \emph{(B) examine thoroughly}, \emph{(C) examine casually}, \emph{(D) check ticket}, \emph{(E) decide}, \emph{(F) re-initiate request}, \emph{(G) pay compensation}, and \emph{(H) reject request}. Figure \ref{sfig:traces_example} shows a small example event log consisting of six execution trails of the process model. The Inductive Miner \cite{Leemans2013} process discovery algorithm provides the guarantee that it can re-discover the process model from an event log given that all pairs of activities that can directly follow each other in the process are present in the event log, i.e., the log is \emph{directly-follows complete}. Since the log in Figure \ref{sfig:traces_example} is directly-follows complete, applying the Inductive Miner to this log results in the process model in Figure \ref{sfig:local_model_example}, which generated the log.

However, the presence of activities that can occur spontaneously at any point in the process execution, which we will call \emph{chaotic activities}, substantially impacts the quality of the resulting process models obtained with process discovery techniques. Figure \ref{sfig:traces_example_noisy} contains the event log obtained from the one in Figure \ref{sfig:traces_example} by adding activity \emph{(X) the customer calls} at random points, since customers can call the call center multiple times at any point in time during the execution of the process. Figure \ref{sfig:local_model_example_noisy} shows the resulting process model discovered by the Inductive Miner \cite{Leemans2013} from the event log of Figure \ref{sfig:traces_example_noisy}. The process model discovered from the ``clean'' example log without activity X (Figure \ref{sfig:local_model_example}) was very simple, interpretable, and accurate with respect to the behavior allowed in the process. In contrast, the process model discovered from the log containing X (Figure \ref{sfig:local_model_example_noisy}) is very complex, hard to interpret, and it overgeneralizes by allowing for too much behavior that is not possible in the process. We consider X to be a so-called \emph{chaotic activity} because it does not have a clear position in the process model and it complicates the discovery of the rest of the process. 
The reason for the decline in the quality of process models discovered from logs with chaotic activities is that the directly follows relations, which many process discovery algorithms operate on, are affected by chaotic activities. Examples of such process discovery algorithms include the Inductive Miner \cite{Leemans2013b}, the Heuristics Miner \cite{Weijters2011}, and Fodina \cite{Fodina2017}. In a sequence of activities $\langle\dots,A,C,\dots\rangle$, where $A$ was directly followed by $C$,  the addition of a chaotic activity $X$ can turn the sequence into $\langle\dots,A,X,C,\dots\rangle$, thereby obfuscating the directly-follows relation between activities A and C.

In this paper, we show that existing approaches do not solve the problem of chaotic activities and we present a technique to handle the issue. This paper is structured as follows: in Section \ref{sec:preliminaries} we introduce basic concepts used throughout the paper. In Section \ref{sec:activity_filtering} we propose an approach to filter out chaotic activities. In Section \ref{sec:evaluation} we evaluate our technique using synthetic data where we artificially insert chaotic activities and check whether the filtering techniques can filter out the inserted chaotic activities. Additionally, Section \ref{sec:evaluation} proposes a methodology to evaluate activity filtering techniques in a real-life setting where there is no ground truth knowledge on which activities are truly chaotic, and motivates this methodology by showing that its results are consistent with the synthetic evaluation on the synthetic datasets. In Section \ref{sec:real_life_evaluation} the results on a collection of seventeen real-life event logs are discussed. In Section \ref{sec:slider} we discuss how the activity filtering techniques can be used in a toggle-based approach for human-in-the-loop process discovery. In Section \ref{sec:related_work} we discuss related techniques in the domains of process discovery and the filtering of event logs. Section \ref{sec:conclusion} concludes this paper and discusses several directions for future work.

\section{Preliminaries}
\label{sec:preliminaries}
In this section, we introduce concepts and notation throughout this paper.

$X=\{a_1,a_2,\dots,a_n\}$ denotes a finite set. $\mathcal{P}(X)$ denotes the power set of $X$, i.e., the set of all possible subsets of $X$. $X{\setminus}Y$ denotes the set of elements that are in set $X$ but not in set $Y$, e.g., $\{a,b,c\}{\setminus}\{a,c\}{=}\{b\}$. $X^*$ denotes the set of all sequences over a set $X$ and $\sigma=\langle a_1,a_2,\dots,a_n\rangle$ denotes a sequence of length $n$, with $\sigma(i)=a_i$ and $\langle\rangle$ the empty sequence. $\sigma{\upharpoonright}_X$ is the projection of $\sigma$ on $X$, e.g. $\langle a,b,c,a,b,c\rangle{\upharpoonright}_{\{a,c\}}=\langle a,c,a,c \rangle$. $\sigma_1\cdot\sigma_2$ denotes the concatenation of sequences $\sigma_1$ and $\sigma_2$, e.g., $\langle a,b,c\rangle\cdot\langle d,e\rangle=\langle a,b,c,d,e\rangle$.

A partial function $f{\in} X {\nrightarrow} Y$ with domain $\mathit{dom}(f)$ can be lifted to sequences over $X$ using the following recursive definition: (1) $f(\langle\rangle)=\langle\rangle$;  (2) for any $\sigma{\in} X^*$ and $x\in X$:
\begin{center}
	$f(\sigma \cdot \langle x\rangle) =
	\left\{
	\begin{array}{ll}
	f(\sigma)  & \mbox{if } x{\notin}\mathit{dom}(f), \\
	f(\sigma) \cdot \langle f(x)\rangle & \mbox{if } x{\in}\mathit{dom}(f).
	\end{array}
	\right.$
\end{center}

A multiset (or bag) over $X$ is a function $B:X{\rightarrow}\mathbb{N}$ which we write as $[a_1^{w_1},a_2^{w_2},\dots,a_n^{w_n}]$, where for $1{\le} i {\le} n$ we have $a_i{\in} X$ and $w_i{\in}\mathbb{N}^{+}$. The set of all multisets over $X$ is denoted $\mathcal{B}(X)$.

In the context of process mining, we assume the set of all \emph{process activities} $\Sigma$ to be given. Event logs consist of sequences of events where each event represents a process activity.
\begin{definition}[Event, Trace, and Event Log]
	An \emph{event} $e$ in an event log is the occurrence of an activity $e{\in}\Sigma$. We call a (non-empty) sequence of events $\sigma{\in}\Sigma^+$ a \emph{trace}. An \emph{event log} $L{\in}\mathcal{B}({\Sigma^+})$ is a multiset of traces.
\end{definition}
$L{=}[\langle a,b,c\rangle^2,\langle b,a,c\rangle^3]$ is an example event log over process activities $\Sigma=\{a,b,c\}$, consisting of 2 occurrences of trace $\langle a,b,c\rangle$ and three occurrences of trace $\langle b,a,c\rangle$. $\mathit{Activities}(L)$ denotes the set of process activities $\Sigma$ that occur in $L$, e.g., $\mathit{Activities}(L) = \{a,b,c\}$. $\#(a,L)$ denotes the number of occurrences of activity $a$ in log $L$, e.g., $\#(a,L)=5$.

A process model notation that is frequently used in the area of process mining is the Petri net. Petri nets can be automatically transformed into process model notations that are commonly used in business environments, such as BPMN and BPEL \cite{Lohmann2009}. A Petri net is a directed bipartite graph consisting of places (depicted as circles) and transitions (depicted as rectangles), connected by arcs. A transition describes an activity, while places represent the enabling conditions of transitions. Labels of transitions indicate the type of activity that they represent. Unlabeled transitions ($\tau$-transitions) represent invisible transitions (depicted as gray rectangles), which are only used for routing purposes and are not recorded in the event log.
\begin{definition}[Labeled Petri net]
	\label{def:lpn}
	A \emph{labeled Petri net} $N=\langle P,T,F,\ell\rangle$ is a tuple where $P$ is a finite set of places, $T$ is a finite set of transitions such that $P{\cap}T{=}\emptyset$,  $F{\subseteq}(P {\times}T){\cup}(T{\times}P)$ is a set of directed arcs, called the flow relation, and $\ell{:}T{\nrightarrow}\Sigma$ is a partial labeling function that assigns a label to a transition, or leaves it unlabeled (the $\tau$-transitions).\looseness=-1
\end{definition}

We write $\bullet{n}$ and $n\bullet$ for the input and output nodes of $n\in P \cup T$ (according to $F$). A state of a Petri net is defined by its \emph{marking} $m{\in} \mathcal{B}(P)$ being a multiset of places. A marking is graphically denoted by putting $m(p)$ tokens on each place $p{\in}P$. State changes occur through transition firings. A transition $t$ is enabled (can fire) in a given marking $m$ if each input place $p{\in}{\bullet}t$ contains at least one token. Once $t$ fires, one token is removed from each input place $p{\in}{\bullet} t$ and one token is added to each output place $p'{\in}t \bullet$, leading to a new marking $m'{=}m{-}\bullet\!{t}+t\bullet$.

A firing of a transition $t$ leading from marking $m$ to marking $m'$ is denoted as step $m {\step{t}} m'$. Steps are lifted to sequences of firing  enabled transitions, written $m {\step{\gamma}} m'$ and $\gamma {\in}T^*$ is a \emph{firing sequence}.\looseness=-1

Defining an \emph{initial} and a set of \emph{final} markings allows defining the \emph{language} accepted by a Petri net as a set of finite sequences of activities.

\begin{definition}[Accepting Petri Net]
	An \emph{accepting Petri net} is a triplet $\APN=(N,m_0,\MF)$, where $N$ is a labeled Petri net, $m_0{\in}\mathcal{B}(P)$ is its initial marking, and $\MF{\subseteq}\mathcal{B}(P)$ is its set of possible final markings. A sequence $\sigma{\in}\Sigma^*$ is a \emph{trace} of an accepting Petri net $\APN$ if there exists a firing sequence $m_0{\step{\gamma}}m_f$ such that $m_f{\in}\MF$, $\gamma{\in}T^*$ and $\ell(\gamma){=}\sigma$.
\end{definition}

In the Petri nets that are shown in this paper, places that belong to the initial marking contain a token and places belonging to a final marking contain a bottom right label $f_i$ with $i$ a final marking identifier or are simply marked as $\begin{tikzpicture}
[node distance=1.4cm,
on grid,>=stealth',
bend angle=20,
auto,
every place/.style= {minimum size=0.1mm},
]
\node [place,pattern=custom north west lines,hatchspread=1.5pt,hatchthickness=0.25pt,hatchcolor=gray] {};
\end{tikzpicture}$ in case of a single final marking.

The \emph{language} $\Lan(\APN)$ is the set of all its traces, i.e., $\Lan(\APN)=\{l(\gamma) |\allowbreak \gamma{\in}T^*{\land}\exists_{m_f{\in}MF}m_0{\step{\gamma}}m_f\}$, which can be of infinite size when $\APN$ contains loops. While we define the language for accepting Petri nets, in theory, $\Lan(M)$ can be defined for any process model $M$ with formal semantics. We denote the universe of process models as $\mathcal{M}$. For each $M{\in}\mathcal{M}$, $\Lan(M)\subseteq\Sigma^+$ is defined.

A process discovery method is a function $\mathit{PD}:\mathcal{B}({\Sigma^+})\rightarrow\mathcal{M}$ that provides a process model for a given event log. The goal is to discover a process model that is a good description of the process from which the event log was obtained, i.e., it should allow for all the behavior that was observed in the event log (called \emph{fitness}) while it should not allow for too much behavior that was not seen in the event log (called \emph{precision}). For an event log $L$, $\tilde{L}{=}\{\sigma{\in}\Sigma^+|L(\sigma){>}0\}$ is the \emph{trace set} of $L$. For example, for log $L{=}[\langle a,b,c\rangle^2,\langle b,a,c\rangle^3]$, $\tilde{L}{=}\{\langle a,b,c\rangle\langle b,a,c\rangle\}$.
For an event log $L$ and a process model $M$, we say that $L$ is \emph{fitting} on $M$ if $\tilde{L}{\subseteq}\Lan(M)$. \emph{Precision} is related to the behavior that is allowed by a model $M$ that was not observed in the event log $L$, i.e., $\Lan(M){\setminus}\tilde{L}$.

\section{Information-Theoretic Approaches to Activity Filtering}
\label{sec:activity_filtering}
We consider a \emph{chaotic activity} to be an activity that can occur at any point in the process and that thereby complicates the discovery of the rest of the process by obfuscating the directly-follows relations of the event log. In this section, we propose a technique to detect chaotic activities in event logs and to filter them out from those event logs.\\

We  extend the function $\#(a,L)$  to the function  $\#(\sigma,L)$ to count the number of occurrence of a sequence $\sigma$, in $L$:\\ $\#(\sigma,L){=}\sum_{\sigma'{\in}L} |\{0{\le} i{\le}{|\sigma'|}{-}{|\sigma|}~\big|~\forall_{1{\le} j{\le}|\sigma|}\sigma'(i{+}j){=}\sigma(j)\}|$.

The \emph{directly-follows ratio}, denoted $\mathit{dfr}(a,b,L)$, represents the ratio of the events of activity $a$ that are directly followed by an event of activity $b$ in event log $L$, i.e., $\mathit{dfr}(a,b,L){=}\frac{\#(\langle a,b\rangle,L)}{\#(a,L)}$.

Likewise, the \emph{directly-precedes ratio}, denoted $\mathit{dpr}(a,b,L)$, represents the ratio of the events of activity $a$ that are directly preceded by an event of activity $b$ in event log $L$, i.e., $\mathit{dpr}(a,b,L){=}\frac{\#(\langle b,a\rangle,L)}{\#(a,L)}$.

$L^\rfloor$ contains the traces of event log $L$ appended with an \emph{artificial end event} that we represent with $\rfloor$. For each $\sigma=\langle e_1,e_2,\dots,e_n\rangle$ in log L, log $L^\rfloor$ contains a trace $\sigma^\rfloor=\langle e_1,e_2,\dots,e_n,\rfloor\rangle$. Likewise, $L^\lfloor$ contains the traces of event log $L$ prepended with an \emph{artificial start event} $\lfloor$, i.e., for each $\sigma=\langle e_1,e_2,\dots,e_n\rangle$ in log L, log $L^\lfloor$ contains a trace $\sigma^\lfloor=\langle \lfloor,e_1,\dots,e_{n}\rangle$. The artificial \emph{start} and \emph{end} events allow us to define the ratio of start events of an activity, e.g., $\mathit{dfr}(a,\rfloor,L^\rfloor)$ and $\mathit{dpr}(a,\lfloor,L^\lfloor)$ represent the ratio of events of activity $a$ that respectively occur at the end of a trace and at the beginning of a trace.

Assuming an arbitrary but consistent order over the set of process activities $\mathit{Activities}(L)$, $\mathit{dfr}(a,L)$ represents the vector of values $\mathit{dfr}(a,b,L^\rfloor)$ for all $b{\in}\mathit{Activities}(L)\cup\{\rfloor\}$ and $\mathit{dpr}(a,L)$ represents the vector of values $\mathit{dpr}(a,b,L^\lfloor)$ for all $b\in\mathit{Activities}(L)\cup\{\lfloor\}$. From a probabilistic point of view, we can regard $\mathit{dfr}(a,L)$ and $\mathit{dpr}(a,L)$ as the empirical estimates of the categorical distributions over respectively the activities directly prior to $a$ and directly after $a$, where the empirical estimates are based on $\#(a,L)$ trials.

\subsection{Direct Entropy-based Activity Filtering}
We define the entropy of an activity in an event log $L$ based on its directly-follows ratio vector and the directly-precedes ratio vector by using the usual definition of function for the categorical probability distribution:
$H(X)=-\sum_{x{\in}X}x\log_2(x)$. We define the entropy of activity $a\in\mathit{Activities}(L)$ in log $L$ as:
$H(a,L)=H(\mathit{dfr}(a,L))+H(\mathit{dpr}(a,L))$. In case there are zero probability values in the directly follows or directly precedes vectors, i.e., $0\in\mathit{dfr}(a,L)\lor0\in\mathit{dpr}(a,L)$, then the value of the corresponding summand $0\log_2(0)$ is taken as $0$, which is consistent with the limit $\lim\limits_{p\to0+}p\log_2(p)=0$.

For example, let event log $L=[\langle a,b,c,x\rangle^{10},\langle a,b,x,c\rangle^{10},\langle a,x,b,c\rangle^{10}]$, then $\mathit{dfr}(a,L)=\langle 0,\frac{20}{30},0,\frac{10}{30},0\rangle$, using the arbitrary but consistent ordering $\langle a,b,c,x,\rfloor\rangle$, indicating that 20 out of 30 events of activity $a$ are followed by $b$ and 10 out of 30 by $x$. Likewise $\mathit{dpr}(a,L){=}\langle 0,0,0,0,1\rangle$, using the arbitrary but consistent ordering $\langle a,b,c,x,\lfloor\rangle$, indicating that all events of activity $a$ are preceded by $\lfloor$. This leads to $H(\mathit{dfr}(a,L))=0.918$, $H(\mathit{dpr}(a,L))=0$, and $H(a,L)=0.918$. Furthermore, $H(b,L)=1.837$, $H(c,L)=1.837$, and $H(x,L)=3.170$, showing that activity $x$ has the highest entropy of the probability distributions for preceding and succeeding activities. We conjecture that activities that are chaotic and behave randomly to a high degree have high values of $H(a,L)$.

\begin{algorithm}[b]
	\caption{An activity filtering approach based on entropy.}
	\label{alg:direct_entropy_greedy}
	\begin{algorithmic}[1]
		\renewcommand{\algorithmicrequire}{\textbf{Input:}}
		\renewcommand{\algorithmicensure}{\textbf{Output:}}
		\REQUIRE event log $L$
		\ENSURE  list of event logs $Q$
		\\ \textit{Initialisation} :
		\STATE $L'=L$
		\STATE $Q=\langle L' \rangle$
		\\ \textit{Main Procedure:}
		\WHILE {$|\mathit{Activities}(L')|>2$}
		\STATE $\mathit{acts}=\mathit{Activities}(L')$
		\STATE $a'=\arg\max_{a\in\mathit{acts}} H(a,L')$
		\STATE $L'=L'\upharpoonright_{\mathit{acts}\setminus\{a'\}}$
		\STATE $Q=Q\cdot \langle L'\rangle$
		\ENDWHILE
		\RETURN $Q$
	\end{algorithmic}
\end{algorithm}

Algorithm \ref{alg:direct_entropy_greedy} describes a greedy approach to iteratively filter the most randomly behaving (chaotic) activity from the event log. The algorithm takes an event log $L$ as input and produces a list of event logs, such that the first element of the list contains a version of $L$ with one activity filtered out, and each following element of the list has one additional activity filtered out compared to the previous element.

In the example event log $L$, Algorithm \ref{alg:direct_entropy_greedy} starts by filtering out activity $x$, followed by activity $b$ or $c$. The algorithm stops when there are two activities left in the event log. The reason not to filter any more activities past this point is closely related to the aim of process discovery: uncovering relations between activities. From an event log with less than two activities no relations between activities can be discovered.

\subsection{The Entropy of Infrequent Activities and Laplace Smoothing}
We defined entropy of the activities in an event log $L$ is based on the directly-follows ratios $\mathit{dfr}$ and the directly-precedes ratios $\mathit{dpr}$ of the activities in $L$. The empirical estimates of the categorical distributions $\mathit{dfr}(a,L)$ and $\mathit{dpr}(a,L)$ become unreliable for small values of $\#(a,L)$. In the extreme case, when $\#(a,L){=}1$, $\mathit{dfr}(a,L)$ assigns an estimate of $1$ to the activity that the single activity $a$ in $L$ happens to be preceded by and contains a probability of $0$ for the other activities. Likewise, when $\#(a,L){=}1$, $\mathit{dpr}(a,L)$ assigns value $1$ to one activity and value $0$ to all others. Therefore, $\#(a,L){=}1$ leads to $H(\mathit{dfr}(a,L)){=}0$ and $H(\mathit{dfr}(a,L)){=}0$. This shows an undesirable consequence of Algorithm \ref{alg:direct_entropy_greedy}, infrequent activities are unlikely to be filtered out. In the extreme case, the activities that occur only once, which are the last in line activities to be filtered out. This effect is undesired, as very infrequent activities should not be the primary focus of the process model discovered from an event log.

We aim to mitigate this effect by applying Laplace smoothing \cite{Zhai2004} to the empirical estimate of the categorical distributions over the preceding and succeeding activities. Therefore, we define a \emph{smoothed version} of the directly-follows and directly-precedes ratios,
$\mathit{dfr}^s(a,b,L){=}\frac{\alpha~+~\#(\langle a,b\rangle,L)}{\alpha({|\mathit{Activities}(L)|+1})+\#(a,L)}$, with smoothing parameter $\alpha{\in}\mathbb{R}_{\ge0}$. The value of $\mathit{dfr}^s(a,b,L)$ will always be between the empirical estimate $\mathit{dfr}(a,b,L)$ and the uniform probability $\frac{1}{|\mathit{Activities}(L)|+1}$, depending on the value $\alpha$. Similar to $\mathit{dfr}$ and $\mathit{dpr}$, $\mathit{dfr}^s(a,L)$ represents the vector of values $\mathit{dfr}^s(a,b,L^\rfloor)$ for all $b{\in}\mathit{Activities}(L)\cup\{\rfloor\}$ and $\mathit{dpr}^s(a,L)$ represents the vector of values $\mathit{dpr}^s(a,b,L^\lfloor)$ for all $b\in\mathit{Activities}(L)\cup\{\lfloor\}$. From a Bayesian point of view, Laplace smoothing corresponds to the expected value of the posterior distribution that consists of the categorical distribution given by $\mathit{dfr}(a,L)$ and a Dirichlet distributed prior that assigns equal probability to each of the possible number of next activities $|\mathit{Activities}(L)|+1$ (including $\rfloor$). Parameter $\alpha$ indicates the weight that is assigned to the prior belief w.r.t. the evidence that is found in the data. An alternative definition of the entropy of log $L$, based on the smoothed distributions over the preceding and succeeding activities, is as follows:
$H^s(a,L)=H(\mathit{dfr}^s(a,L))+H(\mathit{dpr}^s(a,L))$. The smoothed direct entropy-based activity filter is similar to Algorithm \ref{alg:direct_entropy_greedy}, where function $H$ in line 5 of the algorithm is replaced by $H^s$. Function $H(a,L)$ starts from the assumption that an activity is non-chaotic unless we see sufficient evidence in the data for it's chaoticness, function $H^s(a,L)$ in contrast starts from the assumption that is is chaotic, unless we see evidence sufficient evidence in the data for it's non-chaoticness.

Categorical distribution $\mathit{dfr}(a,L)$ consists of $|\mathit{Activities(L)}|+1$, therefore, the maximum entropy of an activity decreases as more activities get filtered out of the event log. The keep the values of $H^s(a,L)$ comparable between iterations of the filtering algorithm, we propose to gradually increase the weight of the prior by setting weight parameter $\alpha$ to $\frac{1}{|\mathit{Activities(L)}|}$.

\subsection{Indirect Entropy-based Activity Filtering}
An alternative approach to the method proposed in Algorithm \ref{alg:direct_entropy_greedy} is to filter out activities such that the \emph{other} activities in the log become less chaotic. We define the total entropy of an event log $L$ as the sum of the entropies of the activities in the log, i.e.,
$H(L)=\sum_{a\in\mathit{Activities}(L)}H(a,L)$.

Algorithm \ref{alg:indirect_entropy_greedy} describes a greedy approach that iteratively filters out the activity that results in the lowest total log entropy. We call this approach the \emph{indirect entropy-based activity filter}, as opposed to the \emph{direct entropy-based activity filter} (Algorithm \ref{alg:direct_entropy_greedy}), which selects the to-be-filtered activity directly based on the activity entropy, instead of based on the total log entropy after removal.
\begin{algorithm}[t]
	\caption{An indirect activity filtering approach based on entropy.}
	\label{alg:indirect_entropy_greedy}
	\begin{algorithmic}[1]
		\renewcommand{\algorithmicrequire}{\textbf{Input:}}
		\renewcommand{\algorithmicensure}{\textbf{Output:}}
		\REQUIRE event log $L$
		\ENSURE  list of event logs $Q$
		\\ \textit{Initialisation} :
		\STATE $L'=L$
		\STATE $Q=\langle L' \rangle$
		\\ \textit{Main Procedure:}
		\WHILE {$|\mathit{Activities}(L')|>2$}
		\STATE $\mathit{acts}=\mathit{Activities}(L')$
		\STATE $a'=\arg\min_{a\in\mathit{acts}} H(L'\upharpoonright_{\mathit{acts}\setminus\{a\}})$
		\STATE $L'=L'\upharpoonright_{\mathit{acts}\setminus\{a'\}}$
		\STATE $Q=Q\cdot \langle L'\rangle$
		\ENDWHILE
		\RETURN $Q$
	\end{algorithmic}
\end{algorithm}

\subsection{An Indirect Entropy-based Activity Filter with Laplace Smoothing}
Just like the direct entropy-based activity filter, the indirect entropy-based activity filter is sensitive to infrequent activities. To deal with this problem, the ideas of the indirect entropy-based activity filtering method and Laplace smoothing can be combined, using the following definition for smoothed log entropy:

$H^s(L)=\sum_{a\in\mathit{Activities}(L)}H^s(a,L)$.

The algorithm for indirect entropy-based activity filtering with Laplace smoothing is identical to Algorithm \ref{alg:indirect_entropy_greedy}, in which function $H$ in line 5 is replaced by function $H^s$.
\section{Evaluation using Synthetic Data}
\label{sec:evaluation}
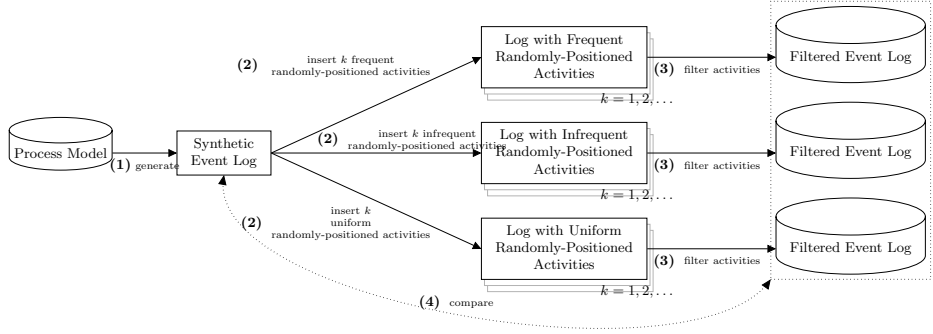
\begin{figure*}[t]
	\centering
	\resizebox{\textwidth}{!}{
		\begin{tikzpicture}[
		dir/.tip = {Triangle[scale length=1.5, scale width=1.2]},
		log/.style = {cylinder,shape border rotate=90,aspect=0.25,minimum width=2cm,minimum height=1cm},
		lpm/.style = {minimum width=2cm,minimum height=1cm,double copy shadow={shadow xshift=0.5ex,shadow yshift=-1ex,draw=black!30},fill=white,draw=black},
		model/.style = {minimum width=2cm,minimum height=1cm}
		]
		\node [draw,log] (llog) {Process Model};
		\node [draw,right = 1.5 of llog] (lpm) {\begin{tabular}{c}Synthetic\\ Event Log\end{tabular}};
		\node [draw,lpm, above right = 1 and 4.9 of lpm,label=below right:{$k=1,2,\dots$}] (flpm) {\begin{tabular}{c}Log with Frequent \\Randomly-Positioned\\ Activities\end{tabular}};
		\node [draw,lpm, right = 4.9 of lpm,label=below right:{$k=1,2,\dots$}] (flpm1) {\begin{tabular}{c}Log with Infrequent \\Randomly-Positioned\\ Activities\end{tabular}};
		\node [draw,lpm, below right = 1 and 4.9 of lpm,label=below right:{$k=1,2,\dots$}] (flpm2) {\begin{tabular}{c}Log with Uniform\\Randomly-Positioned\\ Activities\end{tabular}};
		\node [draw,log, right = 3 of flpm] (hlog) {\begin{tabular}{c}Filtered Event Log\end{tabular}};
		\node [draw,log, right = 3 of flpm1] (hlog1) {\begin{tabular}{c}Filtered Event Log\end{tabular}};
		\node [draw,log, right = 3 of flpm2] (hlog2) {\begin{tabular}{c}Filtered Event Log\end{tabular}};
		\node [inner sep=0pt,outer sep=0pt,minimum size=0,left = 0.5 of hlog] (placeholder) {};
		\node[rectangle,dotted,draw, fit=(hlog) (hlog2)](fit1) {};
		\draw [very thick]
		(llog) edge[-dir] node[below,font=\scriptsize]{{\normalsize\textbf{(1)}} generate} (lpm)
		(lpm.east) edge[-dir] node[above left = 0.5 and -1.6,font=\scriptsize]{{\normalsize\textbf{(2)}} \begin{tabular}{c}insert $k$ frequent\\ randomly-positioned activities\end{tabular}} (flpm.west)
		(lpm.east) edge[-dir] node[above right =-0.1 and -1.5,font=\scriptsize]{{\normalsize\textbf{(2)}}\begin{tabular}{c} insert $k$ infrequent\\ randomly-positioned activities\end{tabular}} (flpm1.west)
		(lpm.east) edge[-dir] node[below left = 0 and -1.6,font=\scriptsize]{{\normalsize\textbf{(2)}}\begin{tabular}{c}insert $k$\\ uniform\\ randomly-positioned activities\end{tabular}} (flpm2.west)
		(flpm.east) edge[-dir] node[below ,font=\scriptsize]{{\normalsize\textbf{(3)}}\begin{tabular}{c}filter activities\end{tabular}} (hlog.west)
		(flpm1.east) edge[-dir] node[below ,font=\scriptsize]{{\normalsize\textbf{(3)}}\begin{tabular}{c}filter activities\end{tabular}} (hlog1.west)
		(flpm2.east) edge[-dir] node[below ,font=\scriptsize]{{\normalsize\textbf{(3)}}\begin{tabular}{c}filter activities\end{tabular}} (hlog2.west)
		(fit1.south west) edge[dotted,arrows=triangle 45 - triangle 45,looseness=0.6,out=-125, in=-100] node[<->,above ,font=\scriptsize]{{\normalsize\textbf{(4)}}\begin{tabular}{c}compare\end{tabular}} (lpm.south)
		;
		\end{tikzpicture}}
	\caption{An overview of the proposed evaluation methodology on synthetic data.}%
	\label{fig:overview_synthetic_evaluation}%
\end{figure*}

\begin{figure}
	\centering
	\subfloat[\label{sfig:marustera12}]{
		\centering
		\includegraphics[width=0.7\linewidth]{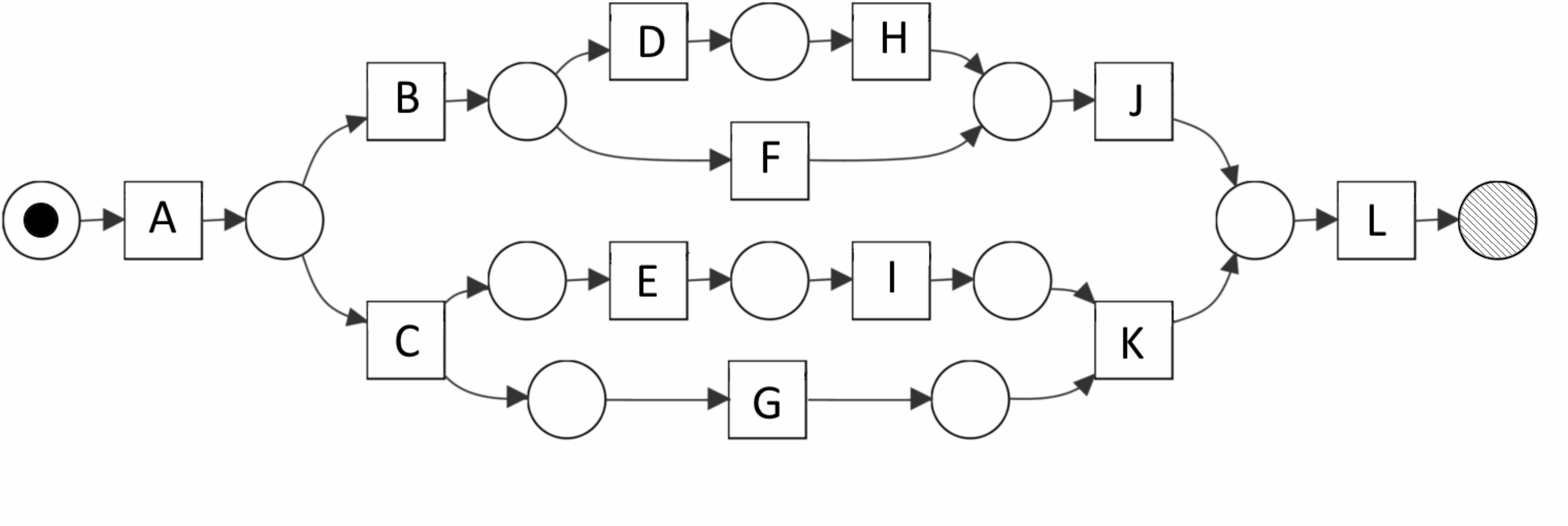}
	}\\
	\subfloat[\label{sfig:marustera12_1}]{
		\centering
		\includegraphics[width=0.9\linewidth]{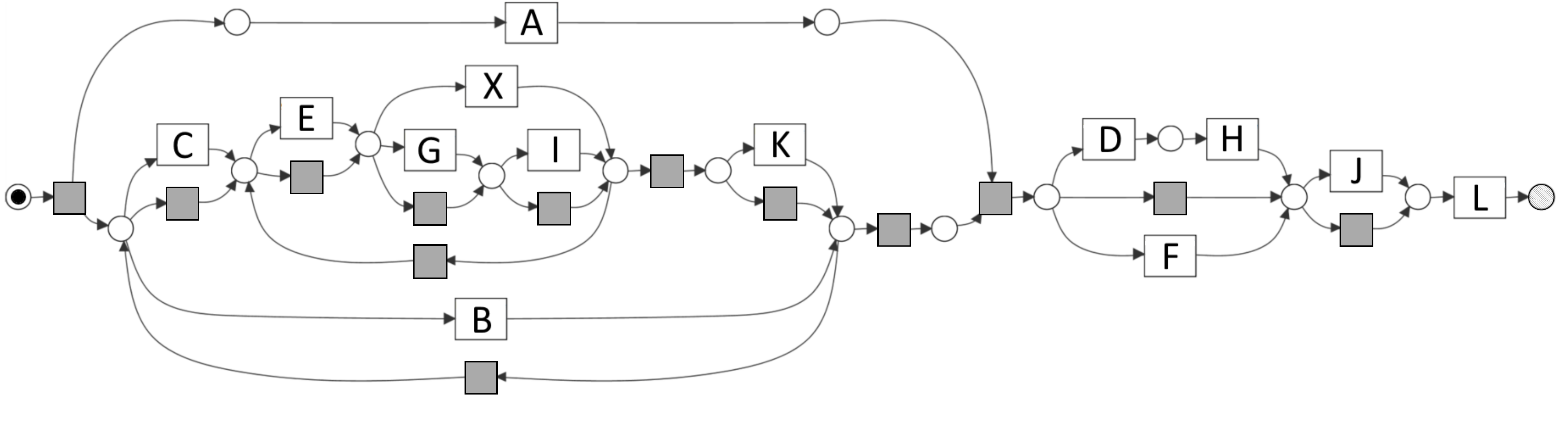}	
	}\\
	\subfloat[\label{sfig:marustera12_2}]{
		\centering
		\includegraphics[width=0.8\linewidth]{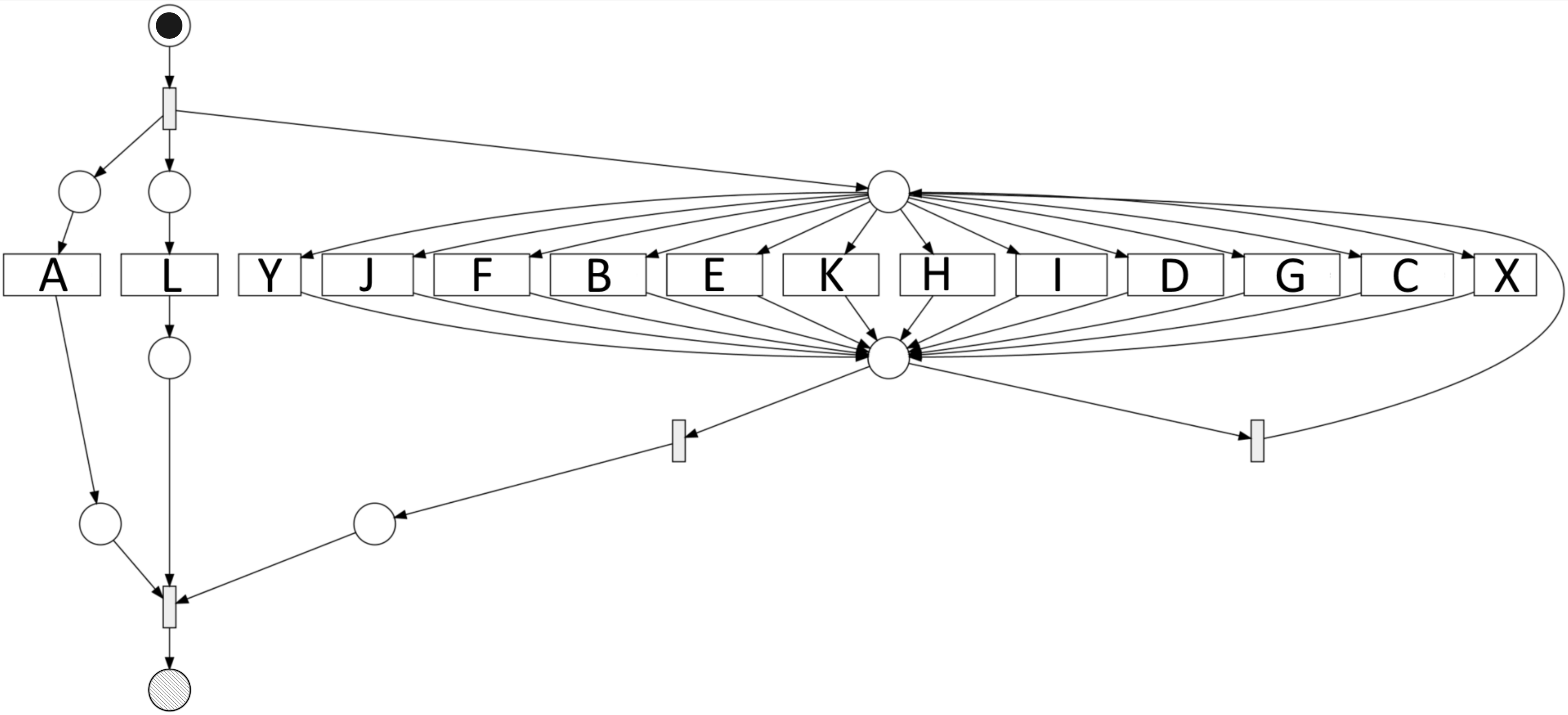}
	}
	\caption{\emph{(a)} The synthetic process model Maruster A12, from which we generate an event log $L$, consisting of 25 traces, from which the process model can be rediscovered with the Inductive Miner \cite{Leemans2013b}, \emph{(b)} the process model discovered by the Inductive Mining when we insert one uniform randomly-positioned activity X to $L_{A12}$, and \emph{(c)} the process model discovered by the Inductive Miner after inserting a second randomly-positioned activity Y to $L_{A12}$.}
\end{figure}

\begin{figure}
	\centering
	\subfloat[\label{sfig:marustera22}]{
		\centering
		\includegraphics[width=0.6\linewidth]{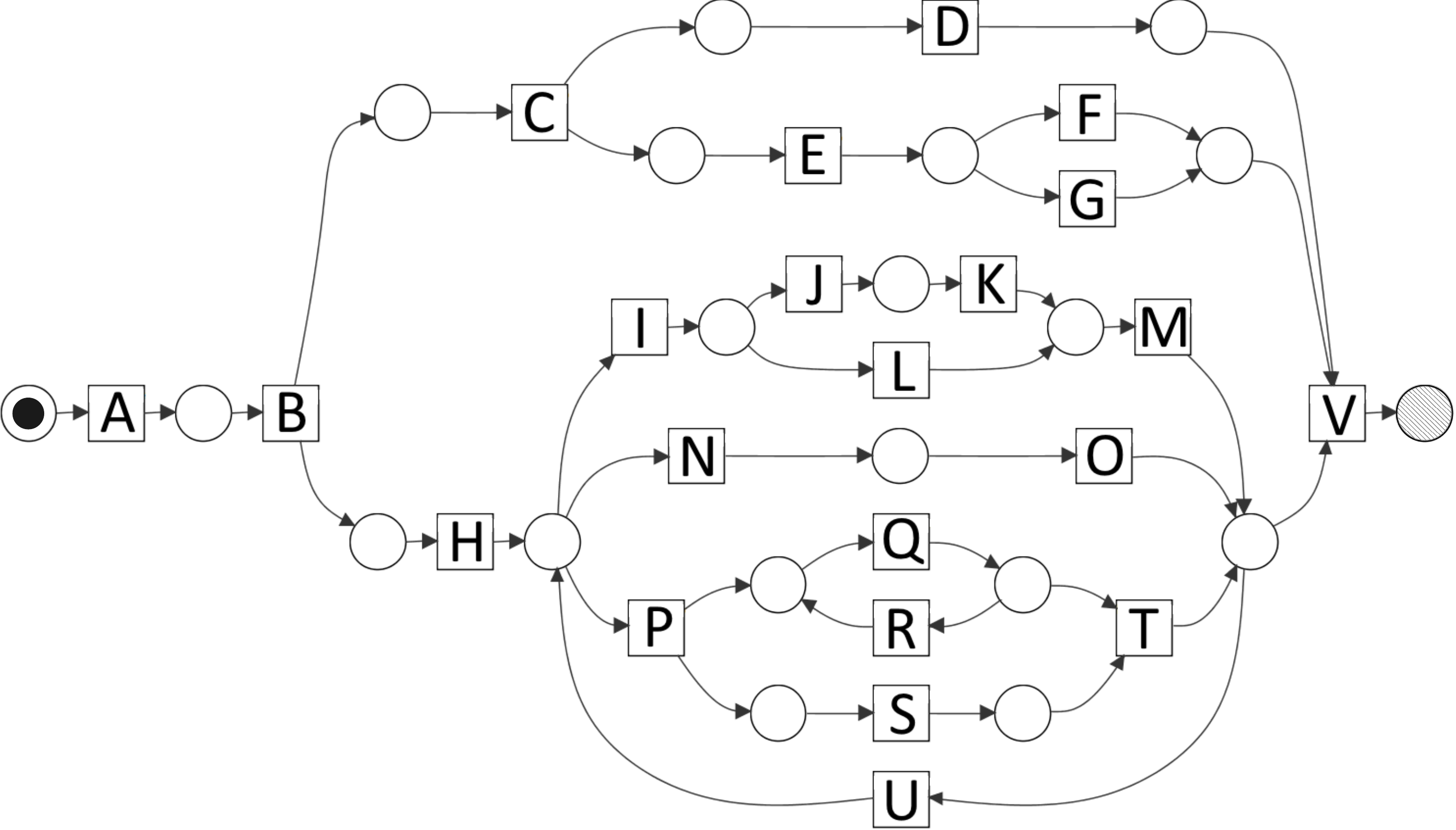}
	}\\
	\subfloat[\label{sfig:marustera22_1}]{
		\centering
		\includegraphics[width=0.9\linewidth]{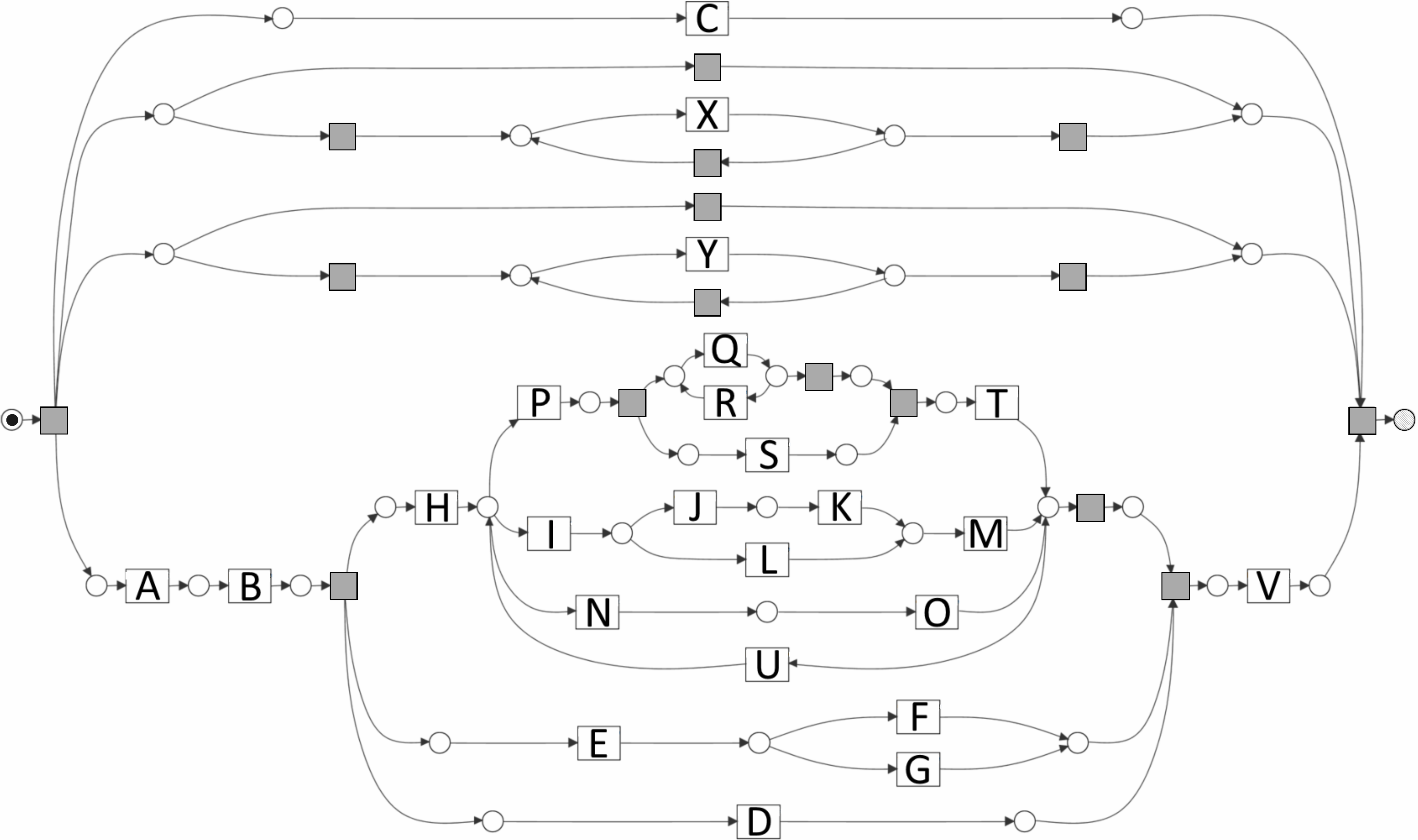}
	}
	\caption{\emph{(a)} The synthetic process model Maruster A22, from which we generate an event log $L_{A22}$, consisting of 400 traces, from which the process model is re-discoverable with the Inductive Miner \cite{Leemans2013b}, and \emph{(b)} the process model discovered by the Inductive Miner after inserting a uniform randomly-positioned activity X to $L_{A22}$.}
\end{figure}

In this section we evaluate the activity filtering techniques using synthetic data. Figure \ref{fig:overview_synthetic_evaluation} gives an overview of the evaluation methodology. First, as step \textbf{(1)}, we generate a synthetic event log from a process model such that we know that all activities of this model are non-chaotic. We take well-known process models introduced by Maruster et al. \cite{Maruster2006}, which respectively consist of 12 and 22 activities and are commonly referred to as the Maruster A12, A22 models. The Maruster A12 and A22 models are shown respectively in Figures \ref{sfig:marustera12} and \ref{sfig:marustera22}. We generated 25 traces by simulation from Maruster A12 to form log $L_{A12}$ and generated 400 traces from Maruster A22 to form log $L_{A22}$. Then, in step \textbf{(2)}, we artificially insert activities that we position at random positions in the log. Since we chose the positions in the log of those activities randomly, we assume those activities to be \emph{chaotic}. We vary the number ($k$) of randomly-positioned activities that we insert, to assess how well the chaotic activity filtering techniques are able to deal with different numbers of randomly-positioned activities in the event log. Furthermore, we vary the frequency of the randomly-positioned activities that we insert, where we distinguish between three types of randomly-positioned activities:
\begin{description}
	\item[\textbf{Frequent randomly-positioned activities}] the number of events inserted for all $k$ randomly-positioned activities is $\max_{a\in\mathit{Activities(L)}}\#(a,L)$.
	\item[\textbf{Infrequent randomly-positioned activities}] the number of events inserted for all $k$ randomly-positioned activities is $\min_{a\in\mathit{Activities(L)}}\#(a,L)$.
	\item[\textbf{Uniform randomly-positioned activities}] for each of the $k$ inserted randomly-positioned activities the frequency is chosen at randomly from a uniform probability distribution with minimum value $\min_{a\in\mathit{Activities(L)}}\#(a,L)$ and maximum value $\max_{a\in\mathit{Activities(L)}}\#(a,L)$.
\end{description}

In step \textbf{(3)} we filter out all the inserted randomly-positioned activities from the event log, by removing activities one-by-one using the activity filtering approaches, until all $k$ artificially inserted activities have been removed again. We then count how many of the activities that were originally in the process model we also removed during this procedure (step \textbf{(4)}). Using this approach, we compare the direct entropy-based activity filtering approach (with and without Laplace smoothing) with the indirect entropy-based activity filtering approach (with and without Laplace smoothing). Furthermore, we compare those activity filtering techniques with activity filtering techniques that are based on the frequency of activities, such as filtering out the activities starting from the least frequent activity (least-frequent-first), or starting from the most frequent activity (most-frequent-first). Frequency-based activity filtering techniques are the current default approach for filtering activities from event logs.

The original process models A12 and A22 can be rediscovered from generated event logs $L_{A12}$ and $L_{A22}$ with the Inductive Miner \cite{Leemans2013b} when there are no added randomly-positioned activities. Figure \ref{sfig:marustera12_1} shows the process model discovered by the Inductive Miner \cite{Leemans2013b} after inserting one uniform randomly-positioned activity, activity $X$, into $L_{A12}$. The insertion of activity $X$ causes the Inductive Miner to create a model that overgeneralizes the behavior of the event log, as indicated by many silent transitions in the process model that allow activities to be skipped. Adding a second uniform randomly-positioned activity $Y$ to $L_{A12}$ results in the Inductive Miner discovering a process model (shown in Figure \ref{sfig:marustera12_2}) that overgeneralizes even further, allowing for almost all sequences over the set of activities. Figure \ref{sfig:marustera22_1} shows the process model discovered by the Inductive Miner after inserting two uniform randomly-paced activities ($X$ and $Y$) into $L_{A22}$. The addition of $X$ and $Y$ has the effect that activity $C$ is no longer positioned at the correct place in the process model, but it is instead put in parallel to the whole process, making the process model overly general, as it wrongly allows for activity $C$ to occur before $A$ and $B$, or after $D$, $E$, $F$, and $G$. Figures \ref{sfig:marustera12_1}, \ref{sfig:marustera12_2} and \ref{sfig:marustera22_1} further motivate the need for filtering out chaotic activities.

Frequent randomly-positioned activities will impact the quality of process models discovered with process discovery to a higher degree than infrequent randomly-positioned activities. Each randomly-positioned activity that is inserted at a random position in the event log is placed in-between two existing events in that log (or at the start or end of the trace). By inserting randomly-positioned activity X in-between two events of activities A and C respectively, the directly-follows relation between activities A and C gets weakened. Therefore, the impact of randomly-positioned activity X is proportional to its frequency $\#(X,L)$.

\subsection{Results}

\begin{table*}
	\centering
	\caption{The number of incorrectly filtered activities per filtering approach on $L_{A12}$ and $L_{A22}$ with $k$ added Uniform (U) / Frequent (F) /Infrequent (I) chaotic activities.}
	\label{tab:maruster_a12_random}
	\scalebox{0.54}{
		\begin{tabular}{|l||c|c|c||c|c|c||c|c|c||c|c|c||c|c|c||c|c|c||c|c|c||c|c|c|}
			\toprule
			&\multicolumn{24}{|c|}{\textbf{Maruster A12} (Number of inserted randomly-positioned activities $\rightarrow$)}\\
			\textbf{Approach}&\multicolumn{3}{|c|}{\textbf{1}} &\multicolumn{3}{|c|}{\textbf{2}} &\multicolumn{3}{|c|}{\textbf{4}} &\multicolumn{3}{|c|}{\textbf{8}} &\multicolumn{3}{|c|}{\textbf{16}} &\multicolumn{3}{|c|}{\textbf{32}} &\multicolumn{3}{|c|}{\textbf{64}} &\multicolumn{3}{|c|}{\textbf{128}} \\
			\midrule
			&U&F&I&U&F&I&U&F&I&U&F&I&U&F&I&U&F&I&U&F&I&U&F&I\\
			\midrule
			Direct & 0&0&0& 0&0&0 & 0&0&0 & 0&0&0 & 0&0&0 & 0&0&12 & 4&0&12 & 10&1&12\\
			Direct ($\alpha{=}\frac{1}{|A|}$)& 0&0&0 & 0&0&0 & 0&0&0 & 0&0&0 & 0&0&0 & 0&0&0 & 4&0&6 & 6&2&12 \\
			Indirect& 0&0&0 & 0&0&0 & 0&0&0 & 0&0&1 & 1&0&1 & 1&0&1 & 2&0&1 & 3&1&6 \\
			Indirect ($\alpha{=}\frac{1}{|A|}$)& 0&0&0 & 0&0&0 & 0&0&0 & 0&0&1 & 1&0&1 & 1&0&1 & 2&0&1 & 2&1&10 \\
			Least-frequent-first & 9&12&0 & 11&12&0 & 6&12&0 & 11&12&0 & 11&12&0 & 12&12&0 & 12&12&0 & 12&12&0 \\
			Most-frequent-first & 11&0&12 & 3&0&12 & 7&0&12 & 10&0&12 & 12&0&12 & 12&0&12 & 12&0&12 & 12&0&12 \\
			\midrule
			&\multicolumn{24}{|c|}{\textbf{Maruster A22} (Number of inserted randomly-positioned activities $\rightarrow$)}\\
			\textbf{Approach}&\multicolumn{3}{|c|}{\textbf{1}} &\multicolumn{3}{|c|}{\textbf{2}} &\multicolumn{3}{|c|}{\textbf{4}} &\multicolumn{3}{|c|}{\textbf{8}} &\multicolumn{3}{|c|}{\textbf{16}} &\multicolumn{3}{|c|}{\textbf{32}} &\multicolumn{3}{|c|}{\textbf{64}} &\multicolumn{3}{|c|}{\textbf{128}} \\
			\midrule
			&U&F&I&U&F&I&U&F&I&U&F&I&U&F&I&U&F&I&U&F&I&U&F&I\\
			\midrule
			Direct & 0&0&0& 0&0&0 &0&0&0 &0&0&1 &0&0&0 &0&0&0 &0&0&0 &0&0&5\\
			Direct ($\alpha{=}\frac{1}{|A|}$)& 0&0&0& 0&0&0 &0&0&0 &0&0&1 &0&0&0 &0&0&0 &0&0&0 &0&0&5 \\
			Indirect& 0&0&0 &0&0&0 & 0&0&0 & 0&0&1 &0&0&1 & 1&0&1 & 1&0&1 & 1&0&1 \\
			Indirect ($\alpha{=}\frac{1}{|A|}$)& 0&0&0 &0&0&0 & 0&0&0 & 0&0&1 &0&0&1 & 1&0&1 & 0&0&1 & 1&0&1 \\
			Least-frequent-first & 16&22&0 & 17&22&0 & 6&22&0 & 21&22&0 & 19&22&0 & 22&22&0 & 22&22&0 & 22&22&0 \\
			Most-frequent-first & 7&0&22 & 8&0&22 & 19&0&22 & 17&0&22 & 19&0&22 & 22&0&22 & 22&0&22 & 22&0&22 \\
			\bottomrule
		\end{tabular}
	}
\end{table*}

Table \ref{tab:maruster_a12_random} reports the number of activities that were originally part of the synthetic process models A12 and A22 that were wrongly filtered out from $L_{A12}$ and $L_{A22}$ as an effect of removing all inserted randomly-positioned activities from these logs. If this number is 12 for Maruster A12 or 22 for Maruster A22 this indicates that all activities of the original process model needed to be filtered out before the activity filtering technique was able to remove all inserted chaotic activities. The results show that the direct filtering approach can perfectly distinguish between actual activities from the process and artificial chaotic activities for up to 32 uniform randomly-positioned activities inserted activities to $L_{A12}$, up to 64 frequent randomly-positioned activities, and up to 16 infrequent randomly-positioned activities. Infrequent randomly-positioned activities are the hardest type of randomly-positioned activities to correctly filter out, as their infrequency can have the effect that the probability distributions over their surrounding activities can \emph{by chance} have low entropy. Using Laplace smoothing with $\alpha=\frac{1}{|Activities(L)|}$ mitigates this effect, but does not completely solve it: the number of incorrectly removed activities drops from 12 to 0 as an effect of Laplace smoothing for 32 added randomly-positioned activities, and from 12 to 6 for 64 added randomly-positioned activities. The indirect activity filter starts making errors already at lower numbers of added randomly-positioned activities than the direct activity filter; however, it is more stable to errors for higher numbers of added randomly-positioned activities, i.e., fewer activities get incorrectly removed for 64 and 128 added randomly-positioned activities. In contrast to direct activity filtering, Laplace smoothing does not seem to reduce the number of wrongly removed activities for indirect activity filtering. In fact, surprisingly, the number of incorrectly removed activities even increased from 6 to 10 as an effect of using Laplace smoothing for 128 infrequent randomly-positioned activities added to $L_{A12}$. The direct and indirect filtering approaches, both with and without Laplace smoothing, outperform the currently widely used approach of filtering out infrequent activities from the event log (least-frequent-first filtering). Furthermore, a second frequency-based activity filtering technique is included in the evaluation in which the most-frequent activities are removed from the event log (most-frequent first filtering). Both Frequency-based filtering approaches are not able to filter out the randomly-positioned activities inserted to $L_{A12}$ and $L_{A22}$, even for small numbers of added randomly-positioned activities.

\subsection{An Evaluation Methodology for Event Data without Ground Truth Information}
In a real-life data evaluation that we perform in the following section, there is no ground truth knowledge on which activities of the process are chaotic. This motivates a more indirect evaluation in which we evaluate the quality of the process model discovered from the event log after filtering out activities with the proposed activity filtering techniques. In this section we propose a methodology for evaluation of activity filtering techniques by assessing the quality of discovered process models, we apply this evaluation methodology to the Maruster A12 and Maruster A22 event logs, and we discuss the agreement between the findings of Table \ref{tab:maruster_a12_random} and the quality of the discovered process models.

There are several ways to quantify the quality of a process model for an event log. Ideally, a process model $M$ should allow for all behavior that was observed in the event log $L$, i.e., $\tilde{L}\setminus\Lan(M)$ should be as small as possible, preferably empty. The \emph{fitness} quality dimension covers this. Furthermore, model $M$ should not allow for too much additional behavior that was \emph{not} seen in the event log, i.e., $\Lan(M)\setminus\tilde{L}$ should be as small as possible. This aspect is called \emph{precision}. For each process model that we discovered, we measure \emph{fitness} and \emph{precision} with respect to the filtered log. Fitness is measured using the alignment-based fitness measure \cite{Adriansyah2011} and we measure precision using negative event precision \cite{Vandenbroucke2013}. Based on the fitness and precision results we also calculate \emph{F-score} \cite{Weerdt2011}, i.e., the harmonic mean between fitness and precision.

Precision is likely to increase by filtering out one or more activities from an event log \emph{independently of which activities are removed from the log}, as a result of two factors. First, precision measures express $\Lan(M)\setminus\tilde{L}$ in terms of the number of activities that are enabled at certain points in the process, w.r.t. the number of activities seen that were actually observed at these points in the process. With the log and model containing fewer activities after filtering, the number of enabled activities is likely to decrease as well. Secondly, activity filtering leads to log $L'$ that contains less behavior than original log $L$ (i.e., $\tilde{L'}$ is smaller than $\tilde{L}$), this makes it easier for process discovery methods to discover a process model with less behavior. These two factors make precision values between event logs with different numbers of activities filtered out incomparable. The degree to which the behavior of filtered log $L'$ decreases w.r.t. an unfiltered log $L$ depends on the activities that are filtered out: when very chaotic activities are filtered from $L$ the behavior decreases much more than when very structured activities are filtered from $L$. One effect of this is that too much behavior in a process model affects the precision of that model more for the log from which the non-chaotic activities are filtered out than for the log from which the chaotic activities are filtered out.

\begin{figure*}
	\centering
	\includegraphics[width=\linewidth]{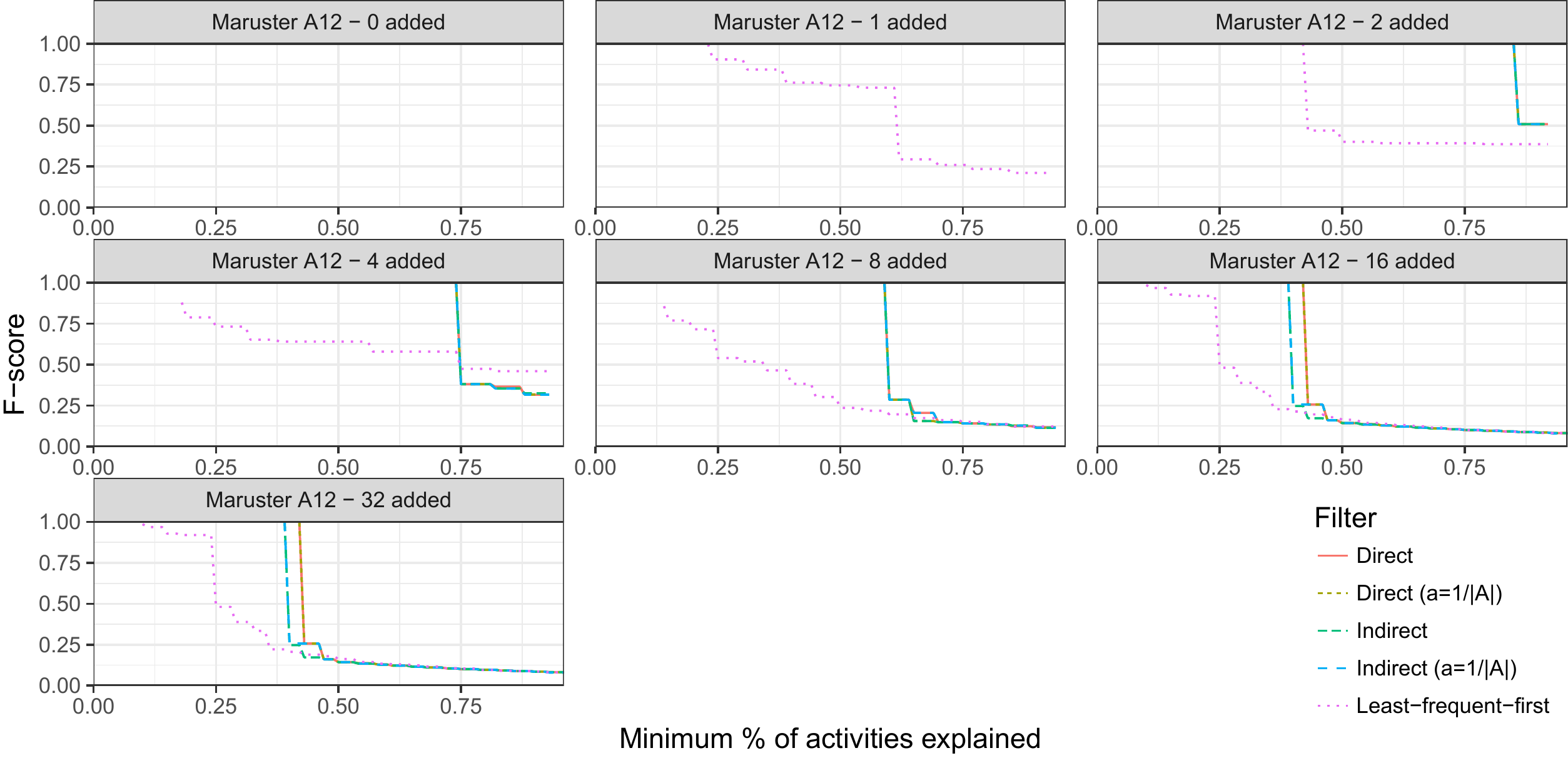}
	\vspace{-0.2cm}
	\caption{F-score on the log generated from the Maruster A12 model with inserted artificial chaotic activities.}
	\label{fig:maruster_logs_fscore}
	\vspace{-0.2cm}
\end{figure*}
\begin{figure*}
	\centering
	\includegraphics[width=\linewidth]{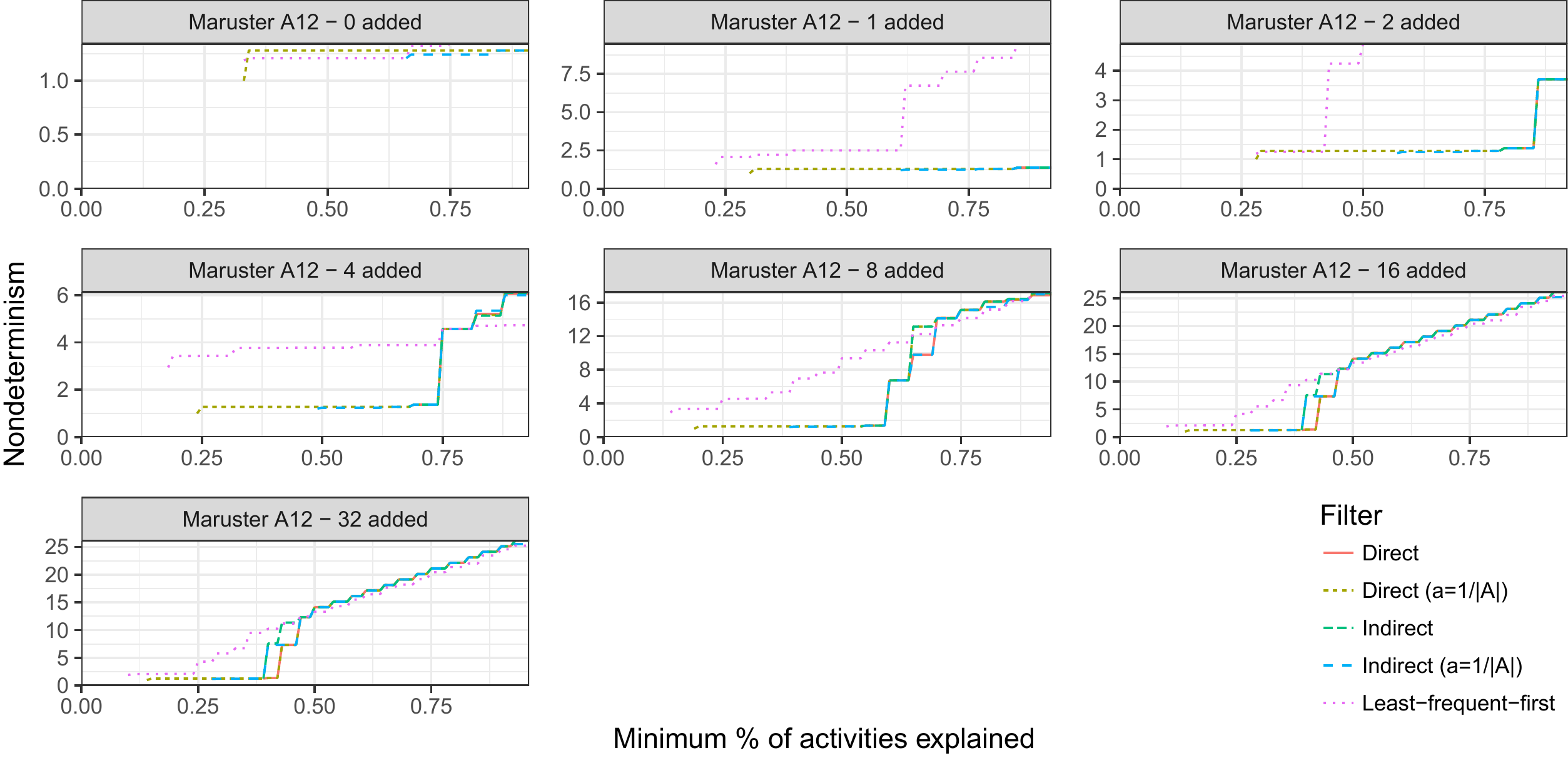}
	\vspace{-0.2cm}
	\caption{Nondeterminism on the log generated from the Maruster A12 model with inserted artificial chaotic activities.}
	\label{fig:maruster_logs_nondeterminism}
	\vspace{-0.2cm}
\end{figure*}

To measure the behavior allowed by the process model independent of which activities are filtered from the event log is to determine the average number of enabled activities when replaying the traces of the log on the model. To deal with traces of the event log that do not fit the behavior of the process model, we calculate \emph{alignments} \cite{Adriansyah2011} between log and model. Alignments are a function $\Gamma^m:\mathcal{M}\times\Sigma^+\rightarrow\mathcal{B}(P)^+$ that map each trace from the event log to a sequence of markings $\langle m_0, \dots, m_f \rangle$ that are reached to replay that trace on the model, with $m_0$ the initial marking and $m_f{\in}\mathit{MF}$, such that for each two consecutive markings $\langle m_i,m_{i+1} \rangle$ there exists a transition $t\in T$ such that $m_{i+1}=m_i-\pre{t}+t\bullet$. Furthermore, alignments also provide a function $\Gamma^t:\mathcal{M}\times\Sigma^+\rightarrow T^+$ that provides the sequence of transitions $\langle t_0,\dots t_n\rangle$ that matches the changes in the sequence of markings, i.e., $m_{1}=m_0-\pre{t_0}+t_0\bullet$, etc. 
For each trace $\sigma\in\Sigma^+$ that fits a process model $N\in\mathcal{M}$ the alignment $l(\Gamma^t(N,\sigma))=\sigma$. For unfitting traces $\sigma\in\Sigma^+$, the alignment is such that $l(\Gamma^t(N,\sigma))$ is as close as possible to $l$ according to some cost function. We refer to Adriansyah et al.~\cite{Adriansyah2011} for a more exhaustive introduction of alignments. Let $\overline{\Gamma^t}$ denote the sequence consisting of only the visible transitions in $\Gamma^t$, and let $\overline{\Gamma^m}$ correspondingly denote the sequence of markings prior to each firing of a visible transition. Given a marking $m\in\mathcal{B}(P)$ we define the nondeterminism of that marking to be the number of reachable visible transitions that can be fired as first next visible transition from $m$, i.e., $\mathit{nondeterminism}(m)=|\{a{\in}\Sigma|m\step{\gamma}m_i\land t{\in}\gamma \land l(t)=a \land \forall_{\gamma_i{\in}\gamma}\gamma_i{\in}\mathit{dom}(l){\implies}\gamma_i{=}t\}|$. We define the nondeterminism of a model $N\in\mathcal{M}$ given a trace $\sigma\in\Sigma^+$ as the average nondeterminism of the markings $\overline{\Gamma^m(N,\sigma)}$ and define the nondeterminism for a model $N$ and a log $L$ as the average nondeterminism over the traces of $L$.

Figure \ref{fig:maruster_logs_fscore} shows the F-scores measured for different percentages of activities filtered out from the Maruster $L_{A12}$ log with different numbers of uniform chaotic activities added. Note that the line stops when further removal of activities does not lead to further improvement in F-score. Note that on the original event log with 0 chaotic activities added the F-score on the original log is already 1.0, resulting in no lines being drawn. With one chaotic activity added, the least-frequent-first filter needs to remove 75\% of the activities before it ends up with F-score 1, which can be explained by the fact that 9 out of 12 non-chaotic needed to be removed in order with the least-frequent-first filter to remove all uniform chaotic activities, as shown in Table \ref{tab:maruster_a12_random}. All entropy-based activity filtering techniques remove the chaotic activity in the first filtering step, immediately leading to an F-score of 1.0. Up until 8 added chaotic activities there is no difference between the entropy-based activity filtering techniques in terms of F-score of the resulting process models, which is consistent with the fact that all these filtering techniques were found to filter without errors for these number of inserted chaotic activities in Table \ref{tab:maruster_a12_random}. For 16 and 32 activities, the direct filtering methods outperform the indirect filtering methods, consistent with the fact that the indirect approach made one filtering error according to the ground truth for these numbers of added chaotic activities. Note that the least-frequent-first filter is outperformed by the entropy-based filtering methods in terms of F-score of the discovered models, as would be expected given the filtering results according to the ground truth.

Figure \ref{fig:maruster_logs_nondeterminism} shows the results in terms of nondeterminism measured for different percentages of activities filtered out from the Maruster $L_{A12}$ log with various numbers of uniform chaotic activities added. The results show very clearly that when filtering out a number of activities that is identical to the number of added chaotic activities (this corresponds to 92\% for one added activity, 86\% for two added activities, 75\% for 4 added activities, 60\% for 8 added activities, 43\% for 16 added activities, and 27\% for 32 added activities), the nondeterminism reaches a value of 1.5, which is the nondeterminism value of the model discovered from the original log without added chaotic activities. The least-frequent-first filter, however, leads to process models where many activities are enabled on average, therefore overgeneralizing the process behavior, as an effect of filtering out nonchaotic activities instead of the added chaotic activities.

\section{Evaluation using Real Life Data}
\label{sec:real_life_evaluation}
\begin{table}
	\centering
	\caption{An overview of the event logs used in the experiments}
	\label{tab:event_logs}
	\resizebox{0.8\linewidth}{!}{
		\begin{tabular}{|l|c|r|r|r|}
			\toprule
			Name & Category & \# traces & \# events & \# activities \\
			\midrule
			BPI'12 \cite{Dongen2012} & Business & 13087 & 164506 & 23\\
			BPI'12 resource 10939 \cite{Tax2016} & Business & 49 & 1682 & 14\\
			Environmental permit \cite{Buijs2014} & Business & 1434 & 8577 & 27\\
			SEPSIS \cite{Mannhardt2016} & Business & 1050 & 15214 & 16 \\
			Traffic Fine \cite{Leoni2015} & Business &150370&561470& 11\\
			Bruno \cite{Bruno2013} & Human behavior & 57 & 553 & 14\\
			CHAD 1600010 \cite{Mccurdy2000}& Human behavior & 26 & 238 & 10\\
			MIT A \cite{Tapia2004} & Human behavior & 16 & 2772 & 27 \\
			MIT B \cite{Tapia2004} & Human behavior & 17 & 1962 & 20 \\
			Ordonez A \cite{Ordonez2013} & Human behavior & 15 & 409 & 12 \\
			van Kasteren \cite{Kasteren2008} & Human behavior & 23 & 220 & 7 \\
			Cook hh102 labour \cite{Cook2013} & Human behavior & 18 & 576 & 18\\
			Cook hh102 weekend \cite{Cook2013} & Human behavior & 18 & 210 & 18\\
			Cook hh104 labour \cite{Cook2013} & Human behavior & 43 & 2100 & 19\\
			Cook hh104 weekend \cite{Cook2013} & Human behavior & 18 & 864 & 19\\
			Cook hh110 labour \cite{Cook2013} & Human behavior & 21 & 695 & 17\\
			Cook hh110 weekend \cite{Cook2013} & Human behavior & 6 & 184 & 14\\
			\bottomrule
	\end{tabular}}
\end{table}




For the experiments on real-life event logs we do not artificially insert  chaotic activities to event logs, but instead filter directly on the activities that are present in these logs. Whether these logs contain chaotic activities that impact process discovery results is not known upfront. Therefore, we apply different activity filtering techniques to these logs and use them to filter out a varying number of activities, after which we assess the quality of the process model that is discovered from these filtered logs. Table \ref{tab:event_logs} gives an overview of the real-life event logs that we use in the experiment. In total, we include five event logs from the business domain. Furthermore, we include twelve event logs that contain events of human behavior, recorded in smart home environments or through wearable devices. Mining process model descriptions of daily life is a novel application of process mining that has recently gained popularity \cite{Dimaggio2016,Leotta2015,Sztyler2015,Tax2017,Tax2016b}. Furthermore, human behavior event data are often challenging for process discovery because of the presence of highly chaotic activities, like \emph{going to the toilet}. We perform the experiments with activity filtering techniques on real-life data with RapidProM \cite{Aalst2017}, which is an extension that adds process mining capabilities to the RapidMiner platform for repeatable scientific workflows.

For each event log, we apply seven different activity filtering techniques for comparison: 1) direct entropy filter without Laplace smoothing, 2) direct entropy filter with Laplace smoothing ($\alpha{=}\frac{1}{|\mathit{Activities(L)|}}$), 3) indirect entropy filter without Laplace smoothing, 4) indirect entropy filter with Laplace smoothing ($\alpha{=}\frac{1}{|\mathit{Activities(L)|}}$), 5) least-frequent-first filtering, 6) most-frequent-first filtering, 7) filtering the activities from the log in a random order. Recall that the activity filtering procedure stops at the point where all but two activities are filtered from the event log because process models that contain just one activity do not communicate any information regarding the relations between activities. For each event log and for each activity filtering approach we discover a process model after each filtering step (i.e., after each removal of an activity). The process discovery step is performed with two process discovery approaches: the Inductive Miner \cite{Leemans2013b}, and the Inductive Miner infrequent (20\%) \cite{Leemans2013}.

\subsubsection{Results on Business Process Event Logs}
\begin{figure*}
	\centering
	\includegraphics[width=\linewidth]{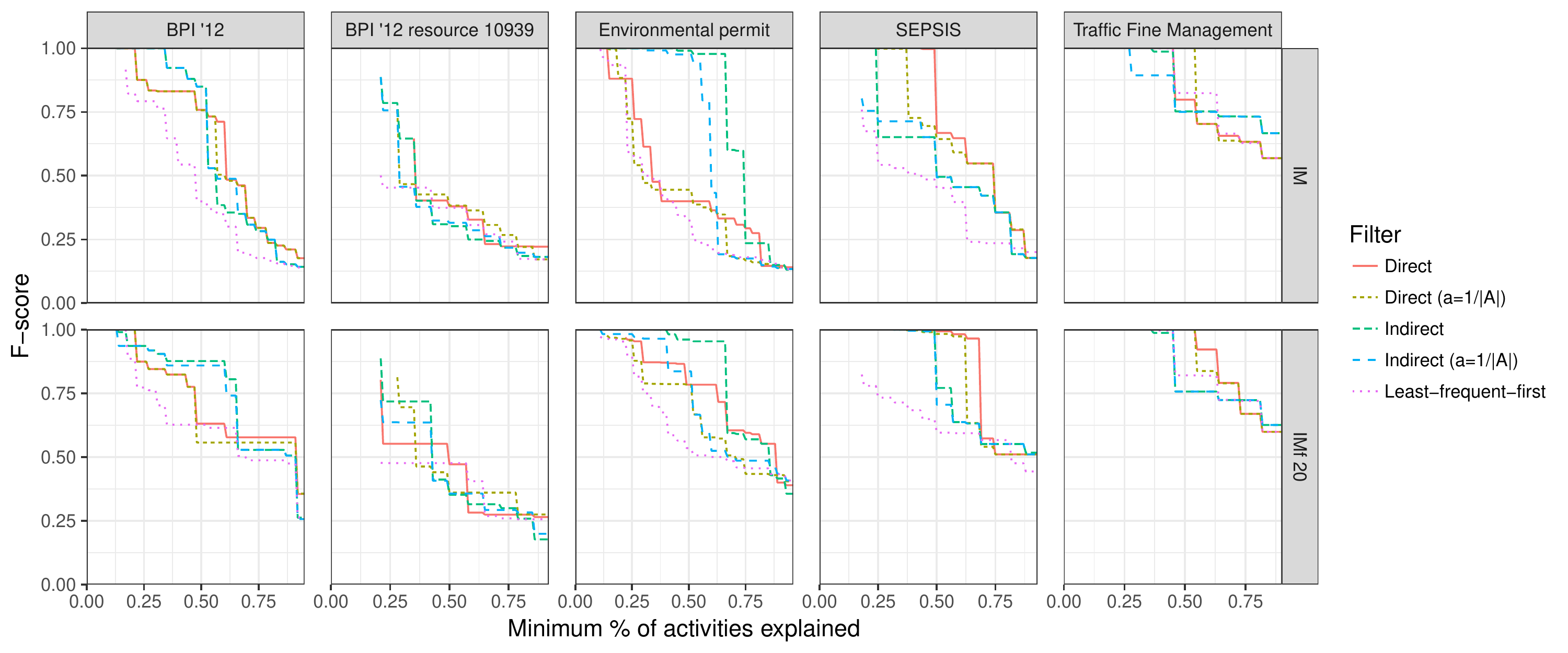}
	\vspace{-0.2cm}
	\caption{F-score on business logs dependent on the minimum share of activities remaining.}
	\label{fig:business_line}
	\vspace{-0.2cm}
\end{figure*}

Figure \ref{fig:business_line} shows the F-score of the process models discovered with the Inductive Miner \cite{Leemans2013b} and the Inductive Miner with infrequent behavior filtering \cite{Leemans2013} (20\% filtering) on the five business event logs for different percentages of activities filtered out and different activity filtering techniques. The figure shows an increasing trend in F-score for all event logs when more activities are filtered from the event log. Furthermore, the line for the least-frequent-first filtering approach is below the lines of the entropy-based filtering techniques for most of the percentages of activities removed on most event logs, which shows that entropy-based filtering enables the discovery of models with higher F-score compared to simply filtering out infrequent activities. There are a few exceptions where filtering out infrequent activities outperforms the entropy-based techniques, e.g., the Inductive Miner on the BPI '12 resource 10939 event log (around 40\% of activities explained) and the traffic fines event log (around 55\% of activities explained). It differs between event logs which of the entropy-based techniques performs best: for the environmental permit log the indirect filter without Laplace smoothing almost dominates the other techniques while for the SEPSIS log the direct filter without Laplace smoothing outperforms the other techniques. Generally, it seems that the use of Laplace smoothing harms F-score, as most parts of the lines of indirect filtering with Laplace smoothing are below the lines of the indirect approach without Laplace smoothing, and similar for the direct approach with and without Laplace smoothing. However, the detrimental effect of Laplace smoothing does not seem to be large, and in some cases, the usage of Laplace smoothing in filtering increases the F-score of the discovered models. 


\begin{figure*}
	\centering
	\includegraphics[width=\linewidth]{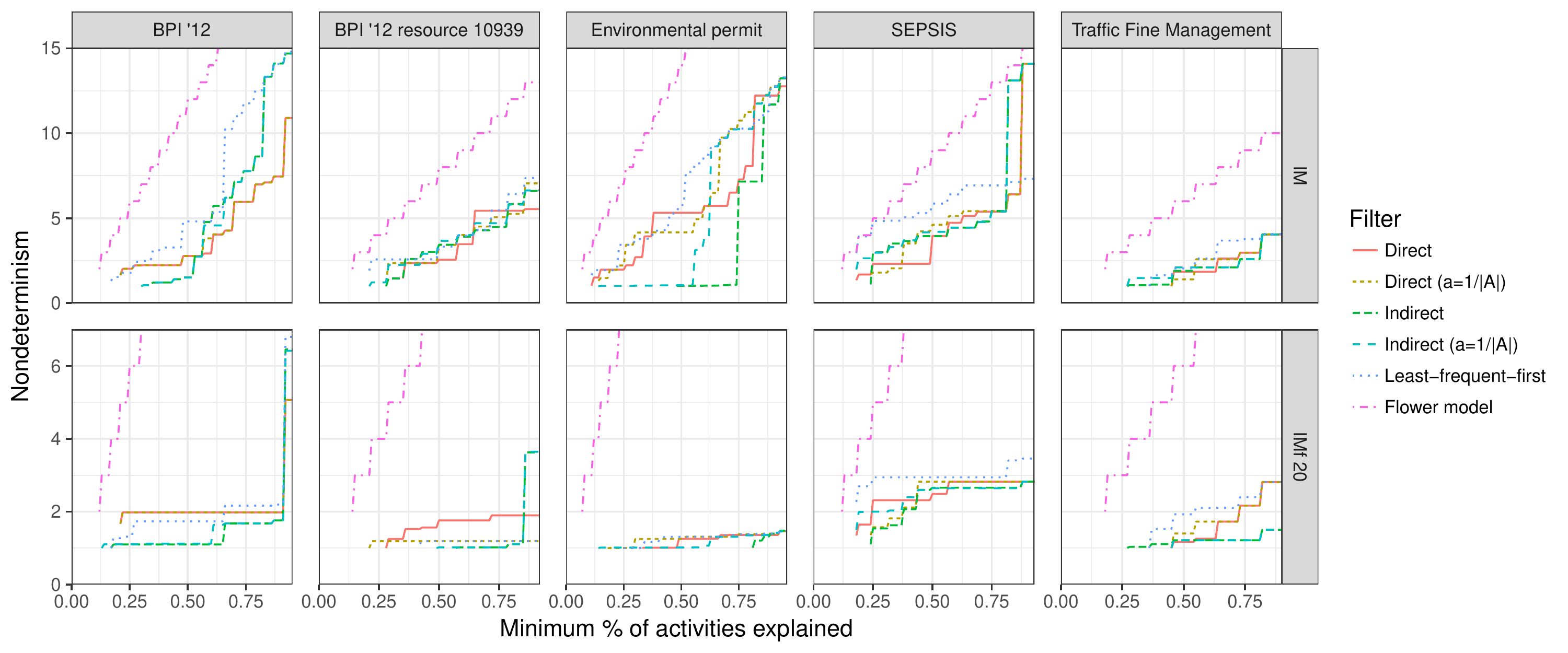}
	\caption{Nondeterminism on business behavior logs dependent on the minimum share of the activities remaining.}
	\label{fig:business_line_nondeterminism}
\end{figure*}

Figure \ref{fig:business_line_nondeterminism} shows the nondeterminism of the process models as a function of the minimum percentage of activities. The green dashed line indicates the nondeterminism of the flower model, i.e., the process model that allows for all behavior over the activities. The lines stop when further removal of activities does not lead to further improvement of nondeterminism. It is clear that the filtering mechanism of the Inductive Miner helps to discover process models that are more behaviorally constrained, as the nondeterminism values are lower for the Inductive Miner infrequent 20\% compared to the Inductive Miner without filtering. However, the results show even when already using the 20\% frequency filter of the Inductive Miner infrequent, the chaotic activity filter can lead to an additional reduction of nondeterminism. Furthermore, the results on the environmental permit log and the SEPSIS log show that filtering several chaotic activities from the event log also enables the discovery of a model with low nondeterminism using the Inductive Miner without filtering. Which of the activity filtering approaches works best seems to be dependent on the event log: the indirect entropy-based filter leads to the models with the lowest nondeterminism on the traffic fine event log, the environmental permit event log, while the direct entropy-based filter works better for some percentages of remaining activities for the SEPSIS log and the BPI '12 resource 10939 log.

\begin{figure}
	\centering
	\includegraphics[width=\linewidth]{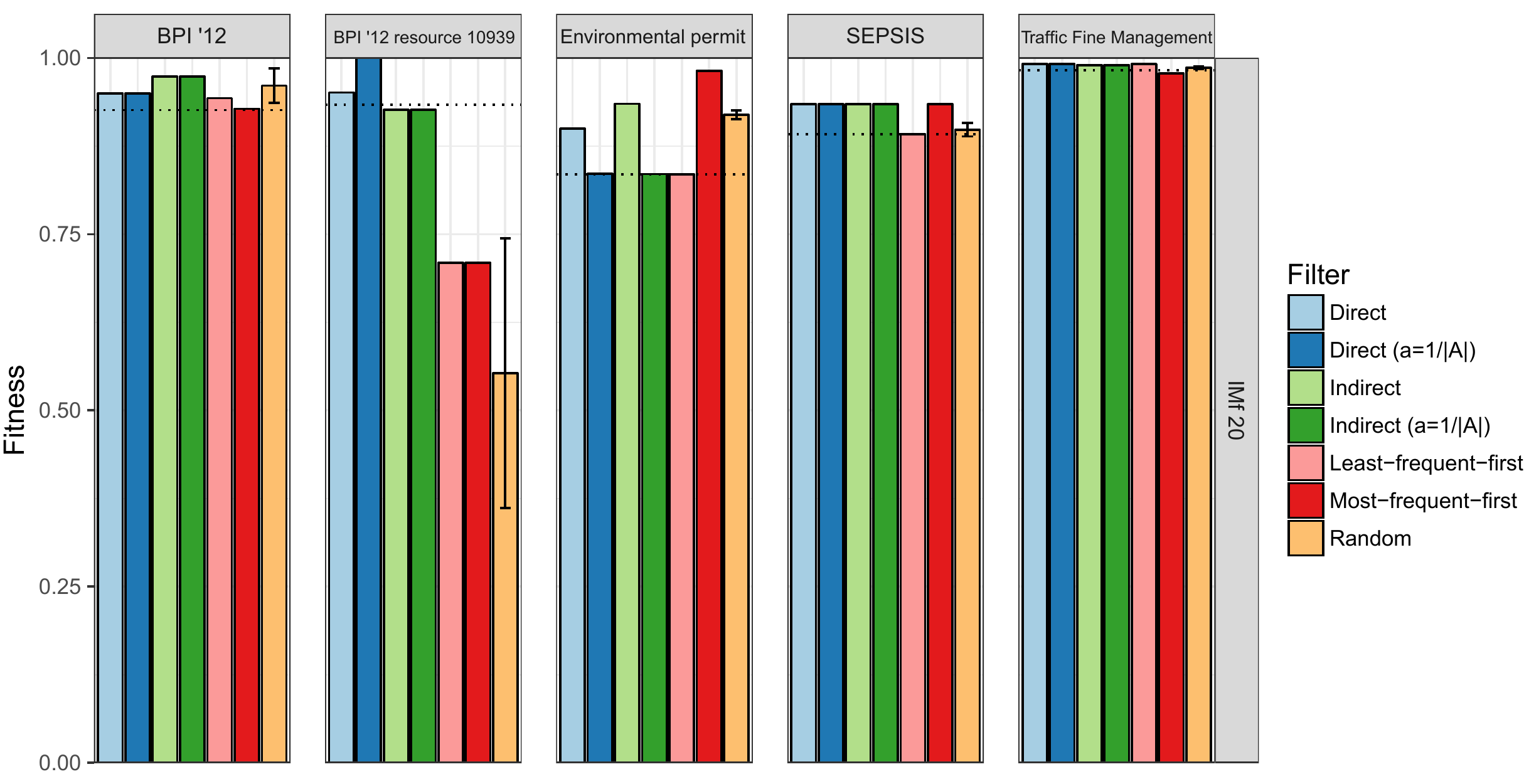}
	\vspace{-0.3cm}
	\caption{Fitness on business logs with least 75\% of the activities remaining.}
	\vspace{-0.4cm}
	\label{fig:business_fitness}
\end{figure}
\begin{figure}
	\centering
	\includegraphics[width=0.9\linewidth]{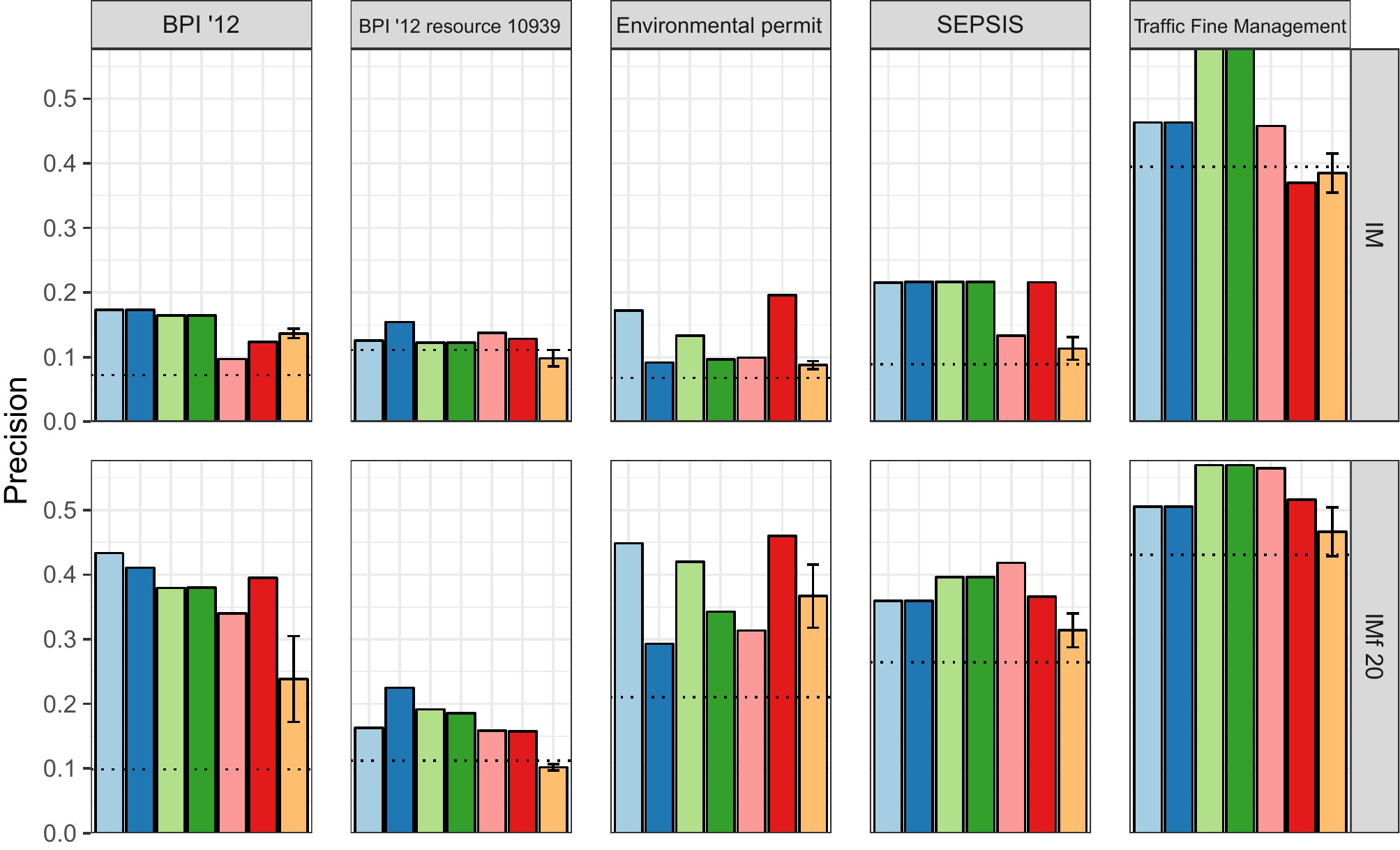}
	\vspace{-0.3cm}
	\caption{Precision on business logs with least 75\% of the activities remaining.}
	\vspace{-0.4cm}
	\label{fig:business_precision}
\end{figure}

Figures \ref{fig:business_fitness} and \ref{fig:business_precision} show the fitness and precision values for the business process event logs at the filtering step that leads to the highest F-score while describing at least 75\% of the activities of the original log. In addition to the filtering techniques shown in Figure \ref{fig:business_line} it also shows the frequency-based activity filter where the \emph{most} frequent activities are filtered out first, and a random baseline is shown which iteratively picks a random activity from the event log to filter out. The error bar for the random activity filter indicates one \emph{standard error of the mean (SEM)} based on eight repetitions of applying the filter. The black dotted horizontal lines indicate the fitness and precision values of the process models discovered from the original event log without filtering any activities. Note that the fitness values are only shown for the Inductive Miner infrequent 20\% \cite{Leemans2013} because the Inductive Miner without infrequent behavior filter \cite{Leemans2013b} provides the formal guarantee that the fitness of the discovered model is $1$. Figure \ref{fig:business_fitness} shows that generally, the differences in fitness between the models discovered from the filtered logs are very minor, and very close to the fitness of the unfiltered log (i.e., the dotted line). Figure \ref{fig:business_precision}, however, shows that the entropy-based filtering approaches outperform filtering out activities based on frequency and filtering out random activities from the event log. The F-scores of the discovered process models is determined mostly by the precision of the models because the activity filtering impacts precision more than it impacts fitness. One exception is the BPI'12 resource 10939 log \cite{Tax2016}, where the fitness decreases to below 0.75 as a result of applying one of the two frequency-based filters, while the precision increase as an effect of applying the filter is only minor.

\begin{figure*}
	\centering
	\includegraphics[width=\linewidth]{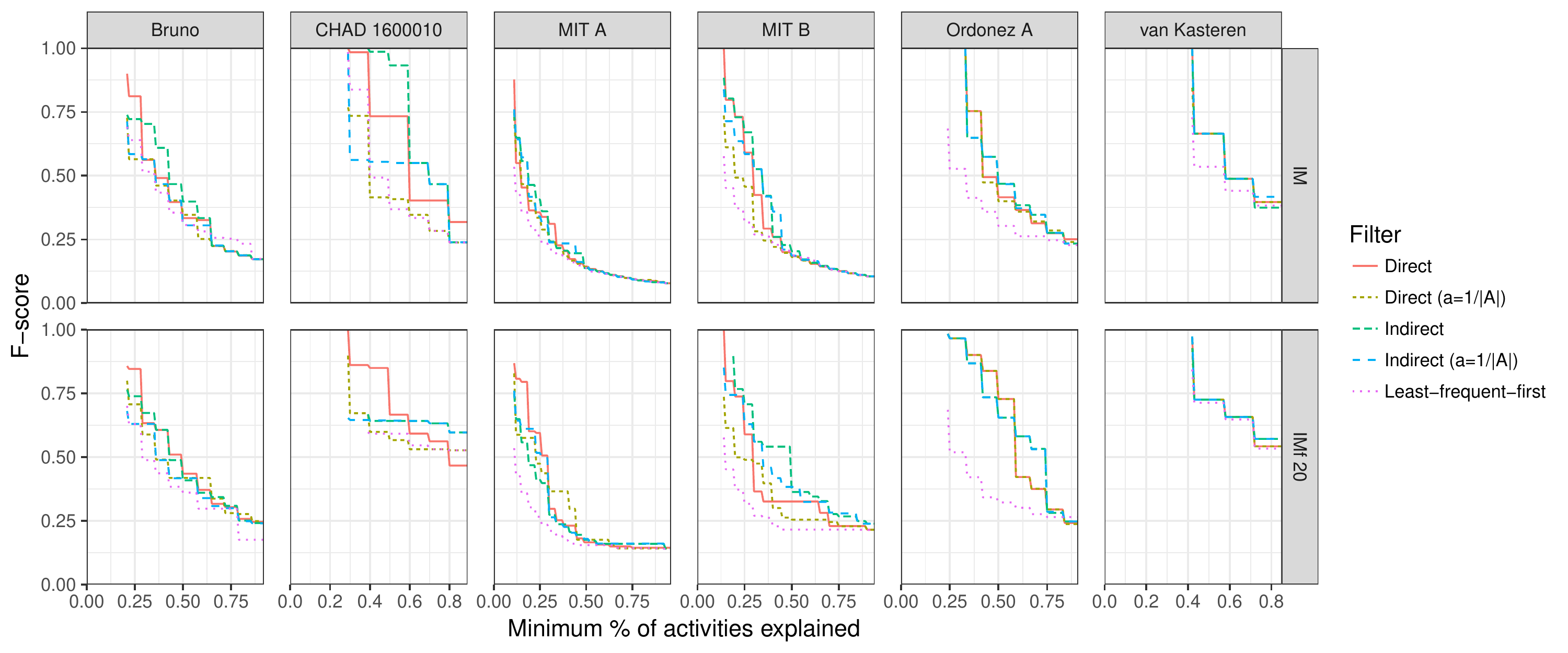}
	\vspace{-0.2cm}
	\caption{F-score on human behavior logs dependent on the minimum share of activities.}
	\vspace{-0.2cm}
	\label{fig:behavior_line}
\end{figure*}

\subsubsection{Results on Human Behavior Event Logs}
Figure \ref{fig:behavior_line} shows the maximum F-score for different human behavior event logs as a function of the minimum percentage of activities that are remaining in the log. Again, the general pattern is that the F-score of the discovered process model decreases when the minimum percentage of events explained increases, as the process discovery task gets easier for smaller numbers of activities. The figure shows that filtering infrequent activities from the event log is dominated in terms of F-score by the entropy-based filtering techniques. Like on the business process event logs, there are mixed results on which of the four configurations of the entropy-based filtering technique leads to the highest F-score: on the CHAD event log the indirect activity filter outperforms the direct activity filter when using the Inductive Miner infrequent 20\%; however, the direct activity filter leads to higher F-score for the Inductive Miner when filtering more than 50\% of the activities.

\begin{figure*}
	\centering
	\includegraphics[width=\linewidth]{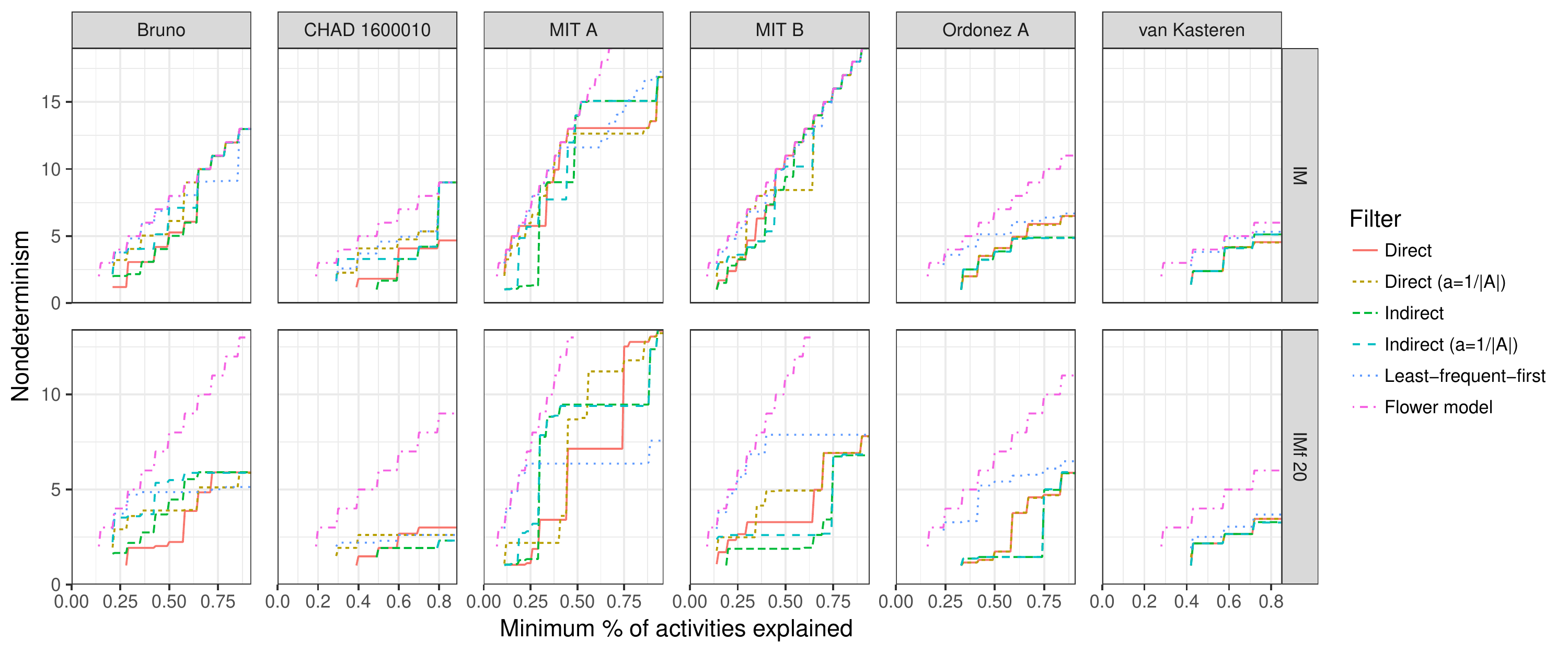}
	\caption{Nondeterminism on human behavior logs dependent on the minimum share of the activities remaining.}
	\label{fig:behavior_line_nondeterminism}
\end{figure*}

Figure \ref{fig:behavior_line_nondeterminism} shows the nondeterminism results for the human behavior event logs. It is noticeable that the nondeterminism values of the process models that are discovered when filtering very few activities are much closer to the flower model compared to what we have seen before for the business process event logs. This is caused by human behavior event logs having much more variability in behavior compared to execution data from business processes, resulting in a much harder process discovery task. After filtering several chaotic activities, the nondeterminism drops significantly to ranges comparable to nondeterminism values seen for logs from the business process domain. This shows that the problem of chaotic activities is much more prominent in human behavior event logs than in business process event logs. The entropy-based activity filtering approaches lead to more deterministic process models compared to filtering out infrequent activities. Two clear examples of this are the MIT B log and the Ordonez A log, on which filtering out infrequent activities after several filtering steps results in a flower model (i.e., nondeterminism is identical to that of the flower model), while entropy-based activity filters enable the discovery of a model with nondeterminism close to one (i.e., very close to a sequential model) while at the same time keeping 75\% of the activities in the event log.

\begin{figure}
	\centering
	\includegraphics[width=1.03\linewidth]{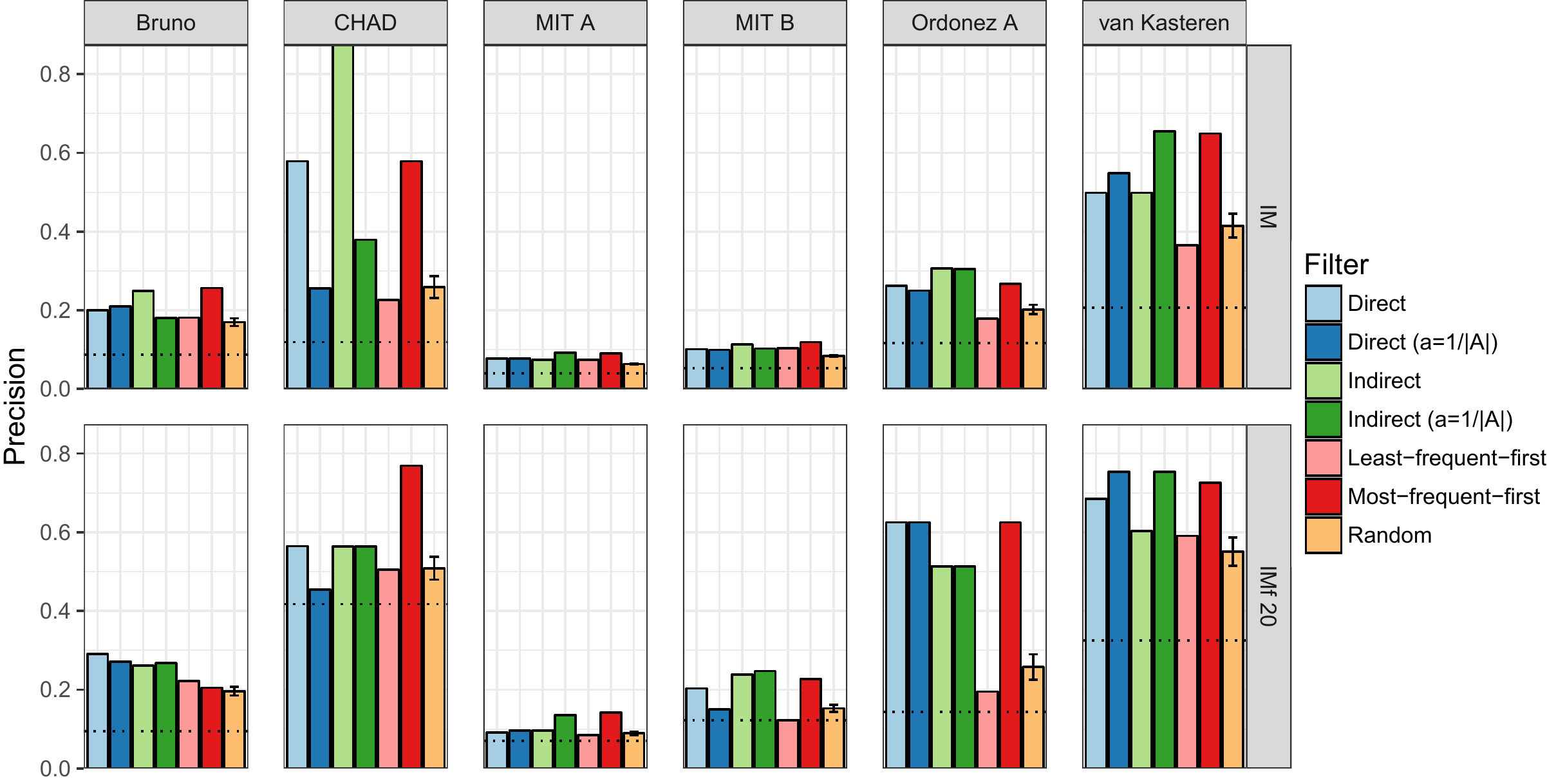}
	\vspace{-0.2cm}
	\caption{Precision on human behavior logs with at least 50\% of the activities.}
	\vspace{-0.5cm}
	\label{fig:behavior_precision}
\end{figure}

Figure \ref{fig:behavior_precision} shows the precision values for the human behavior logs for the filtering step that leads to the highest F-score while describing at least 50\% of the activities of the original log. Similarly to what we have seen in the nondeterminism graph, removing random activities from the log and removing infrequent activities from the log results in smaller precision increases compared to the entropy-based activity filters. Furthermore, it is noticeable that removing frequent activities from the log works quite well to improve the precision of models discovered from the human behavior application domain. The reason for this is that some of the chaotic activities that are present in many of those event logs, including \emph{going to the toilet} and \emph{getting a drink}, also happen to be frequent. On the van Kasteren event log the indirect activity filter with Laplace smoothing leads to the largest increase in precision when mining a model with at least 50\% of the activities (from $0.324$ to $0.732$ with the Inductive Miner infrequent 20\%).

\begin{table}
	\centering
	\caption{Left: the order in which activities are filtered using the direct activity filter with Laplace smoothing ($\alpha=\frac{1}{|Activities(L)|}$) on the van Kasteren log. Right: the order in which the activities are filtered using the least-frequent-first filter.}
	\label{tab:kasteren_removal}
	\scalebox{0.8}{
		\begin{tabular}{|l||c|c|}
			\toprule
			Order & \multicolumn{1}{l|}{\parbox{4.3cm}{\centering Filtered activity \\(indirect entropy-based filter \\with Laplace smoothing)}}& \multicolumn{1}{l|}{\parbox{4.3cm}{\centering Filtered activity\\ (least-frequent-first filter)}}\\
			\midrule
			1 & Use toilet & Prepare dinner\\
			2 & Get drink & Get drink\\
			3 & Leave house & Prepare breakfast\\
			4 & Take shower & Take shower\\
			5 & Go to bed & Go to bed\\
			6 & Prepare breakfast & Leave house\\
			7 & Prepare dinner & Use toilet\\
			\bottomrule
	\end{tabular}}
\end{table}

\begin{figure}
	\centering
	\subfloat[\label{sfig:kasteren_indirect_entropy_laplace}]{\includegraphics[width=0.8\linewidth]{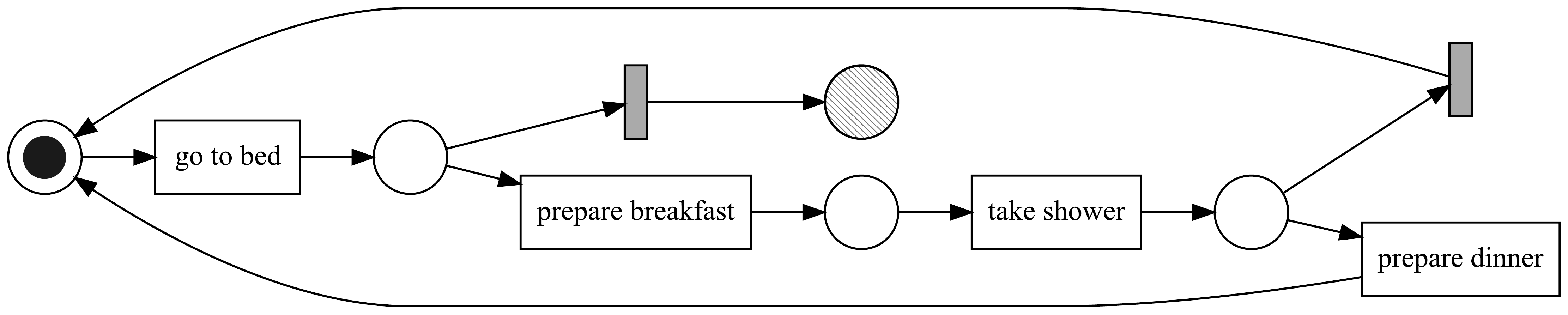}}\\
	\subfloat[\label{sfig:kasteren_infrequent}]{\includegraphics[width=0.8\linewidth]{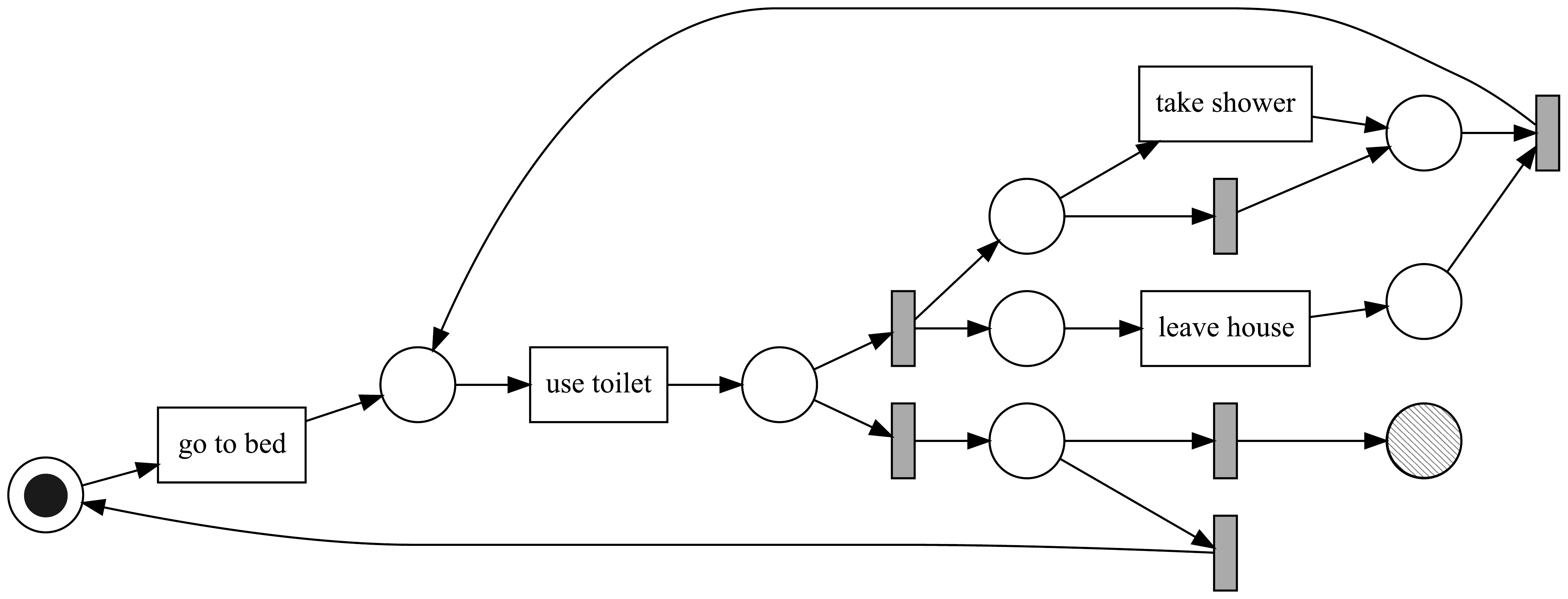}}
	\caption{\emph{(a)} The model discovered with Inductive Miner infrequent 20\% on the Van Kasteren log after filtering all but four activities with the indirect approach with Laplace smoothing, and \emph{(b)} the model discovered from the same log with the same miner when filtering all but four activities when filtering out the least frequent activities.}
\end{figure}

Table \ref{tab:kasteren_removal} shows in which order activities are filtered from the van Kasteren event log by 1) the indirect entropy-based activity filter with Laplace smoothing and 2) the least-frequent-first filter. It shows that the entropy-based filter filters \emph{use toilet} as the first activity, which from domain knowledge we know to be a chaotic activity, as people generally just go to the toilet whenever they need to, regardless of which other activities they have just performed. For the infrequent activity filter \emph{use toilet} would be the last choice of the activities to filter out, because it is the most frequent activity in the van Kasteren event log.

Figures \ref{sfig:kasteren_indirect_entropy_laplace} and \ref{sfig:kasteren_infrequent} show the corresponding process models discovered with the Inductive Miner infrequent 20\% from the logs filtered with the indirect activity filter with Laplace smoothing and the infrequent activity filter respectively. The process model discovered after filtering three activities with the Indirect entropy-based activity filter with Laplace smoothing is very specific on the behavior that it described: after \emph{going to bed}, either the logging ends, or \emph{prepare breakfast} occurs next, followed by \emph{taking a shower}. After taking a shower, there is a possibility to either \emph{go to bed} again or to \emph{prepare dinner} before \emph{going to bed}. The process model discovered after filtering three activities with the infrequent activity filter allows for many more traces: it starts with \emph{go to bed} followed by \emph{use toilet}, after which any of the activities \emph{go to bed}, \emph{take shower}, and \emph{leave house} can occur as next event or the logging can end. Furthermore, the activities \emph{leave house} and \emph{take shower} can occur in any order, and \emph{take shower} can also be skipped.

\begin{figure*}
	\centering
	\includegraphics[width=\linewidth]{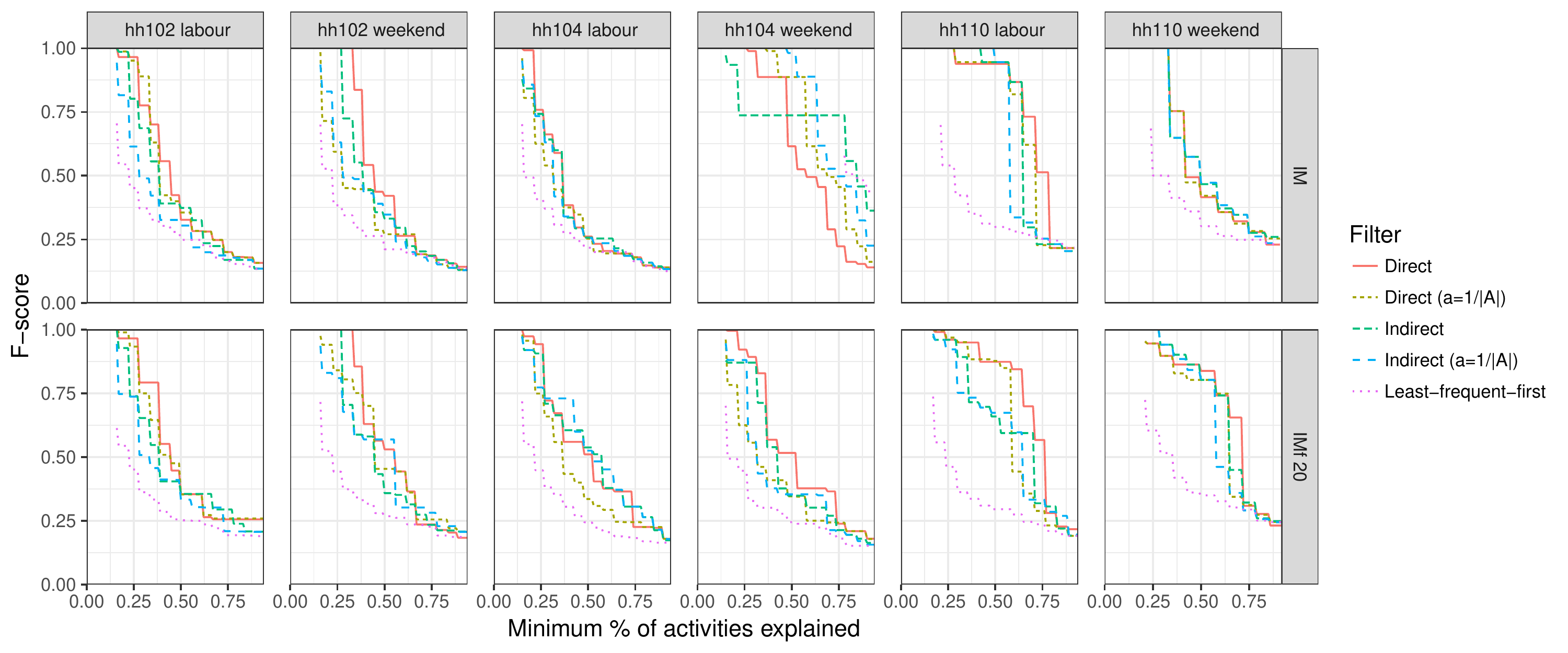}
	\caption{F-score on cook's human behavior logs dependent on the minimum share of the activities remaining.}
	\label{fig:cook_line}
\end{figure*}
\begin{figure*}
	\centering
	\includegraphics[width=\linewidth]{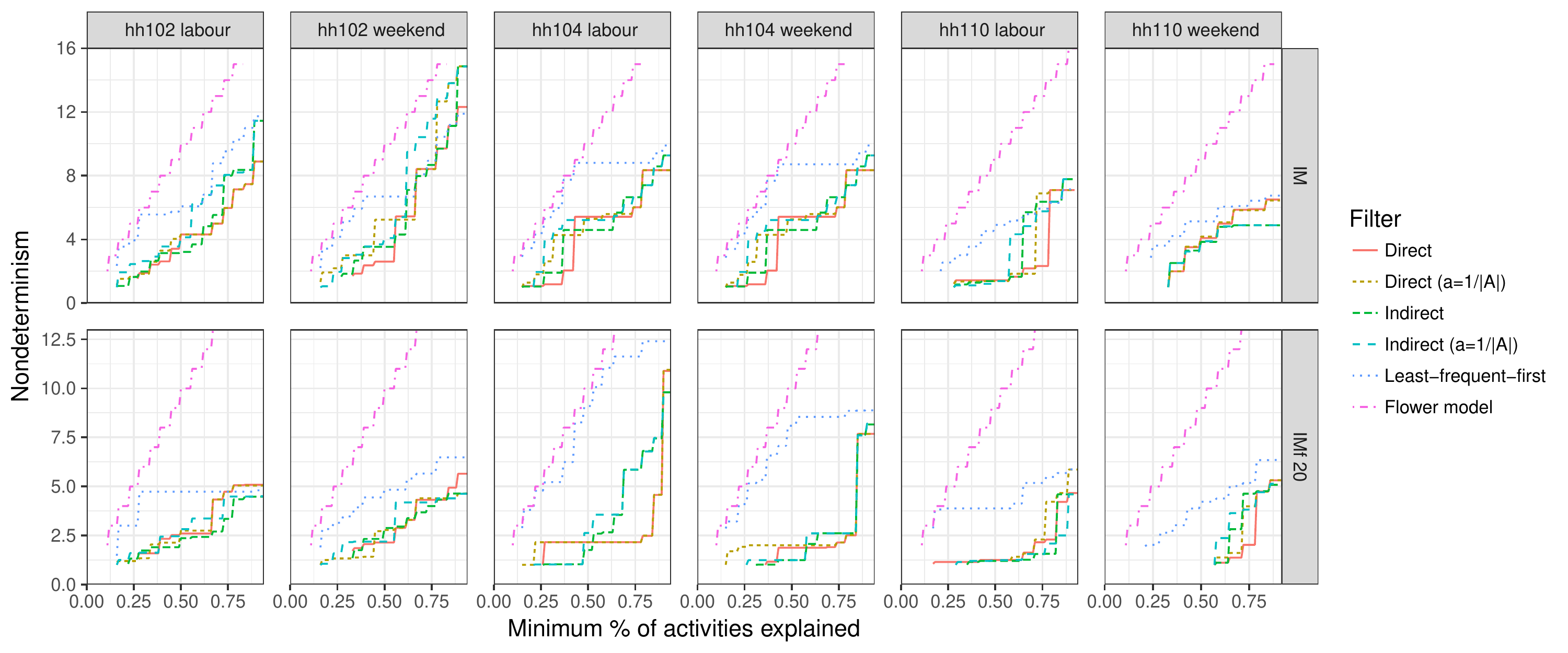}
	\caption{Nondeterminism on cook's human behavior logs dependent on the minimum share of the activities remaining.}
	\label{fig:cook_line_nondeterminism}
\end{figure*}

Figure \ref{fig:cook_line} shows the results on F-score for the human behavior event logs by Cook et al.~\cite{Cook2013}. The results on the Cook event logs are in-line with the results on the human behavior event logs, however, on these event logs, it is even more clear that filtering out infrequent activities leads to suboptimal process models in terms of F-score. Which of the filtering approaches results in the optimal process model in terms of F-score is very dependent on the event log and the minimum number of activities to be remained after filtering: each of the four configurations of the entropy-based filtering approach is optimal for at least one combination of log and minimum percentage of activities explained.

Figure \ref{fig:cook_line_nondeterminism} shows the results in terms of nondeterminism for the same event logs. Filtering infrequent activities at high percentages of activities explained has much lower nondeterminism compared to the flower model, while further left on the graph, after filtering out more activities, the nondeterminism of filtering out infrequent activities gets closer to the flower model. This shows that filtering out infrequent activities can even be harmful to the quality of the obtained process discovery result. The nondeterminism values obtained with the four configurations of the entropy-based filtering approach are generally close to each other, where the optimal configuration is dependent on the log and the number of filtered activities.

\subsubsection{Aggregated Analysis Over All Event Logs}
\begin{figure*}
	\centering
	\includegraphics[width=\linewidth]{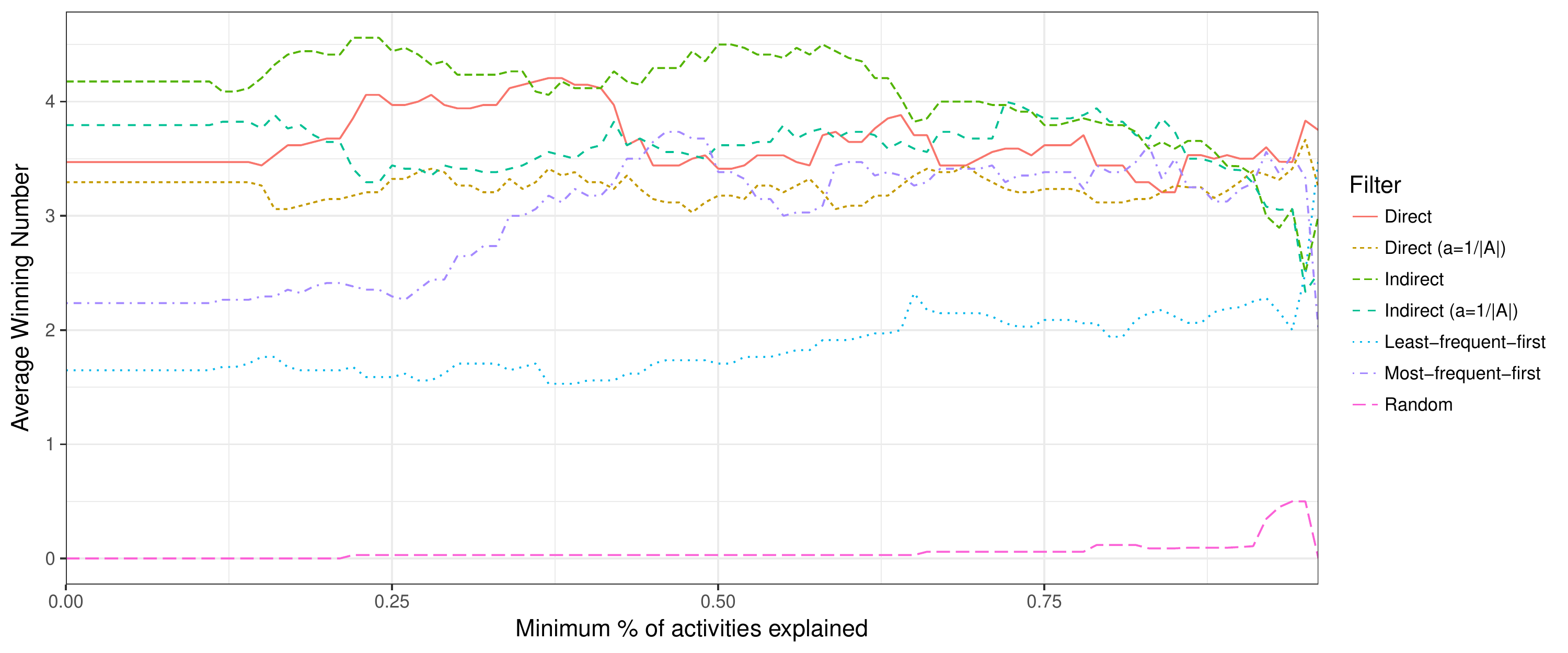}
	\caption{The average winning number for the seven activity filtering techniques dependent on the minimum ratio of activities explained, averaged over the 17 event logs used in the experiment.}
	\label{fig:average_winning_number}
\end{figure*}
\begin{table*}
	\centering
	\caption{Kendall $\tau_b$ rank correlation between five activity filtering methods, mean and standard deviation over the 17 event logs.}
	\resizebox{\textwidth}{!}{
		\begin{tabular}{l|ccccc}
			\toprule
			& Direct & Direct ($\alpha{=}\frac{1}{|A|}$) & Indirect & Indirect ($\alpha{=}\frac{1}{|A|}$) & Least-frequent-first\\
			\midrule
			Direct & 1.0 & 0.2956 & 0.0829 & 0.1408 & 0.0504\\
			Direct ($\alpha{=}\frac{1}{|A|}$) & 0.2956 & 1.0 & 0.0698 & 0.0536 & 0.1454\\
			Indirect & 0.0829 & 0.0698 & 1.0 & 0.6852 & -0.0275\\
			Indirect ($\alpha{=}\frac{1}{|A|}$) & 0.1408 & 0.0536 & 0.6852 & 1.0 & -0.0392\\	
			Least-frequent-first & 0.0504 & 0.1454 & -0.0275 & -0.0392 & 1.0\\
			\bottomrule
	\end{tabular}}
	\label{tab:kendall_correlation_between_filters}
\end{table*}

We have observed in Figures \ref{fig:business_line_nondeterminism}, \ref{fig:behavior_line_nondeterminism}, and \ref{fig:cook_line_nondeterminism} that the entropy-based activity filtering techniques perform differently on different datasets and for different numbers of activities filtered. To evaluate the overall performance of activity filtering techniques, we use the number of other filtering techniques that it can beat over all the seventeen event logs of Table \ref{tab:event_logs}. This metric, known as \emph{winning number}, is commonly used for evaluation in the Information Retrieval (IR) field \cite{Qin2010,Tax2015}. Formally, winning number is defined as

$W^x_i=\sum_{j=1}^{17}\sum_{k=1}^{7} \mathds{1}_{\{N^x_i(j)<N^x_k(j)\}}$

where $j$ is the index of an event log, $i$ and $k$ are indices of activity filtering techniques, $N^x_i(j)$ is the performance of the $i$-th algorithm on the $j$-th event log in terms of nondeterminism where each least $x$\% of activities are explained and $\mathds{1}_{\{N^x_i(j)<N^x_k(j)\}}$ is the indicator function

$\mathds{1}_{\{N^x_i(j)<N^x_k(j)\}}=
\begin{cases}
1,& \text{if } N^x_i(j)<N^x_k(j),\\
0,              & \text{otherwise.}
\end{cases}$

We define $\overline{W}_i^x=\frac{W_i^x}{17}$ as the average number of other activity filtering techniques that are outperformed by filtering technique $i$ at the point where at least $x$\% of activities are explained.


\begin{table*}
	\centering
	\caption{Number of event logs for which we can reject the null hypothesis that the orderings of activities returned by activity filters are uncorrelated, according to the tau test.}
	\resizebox{\textwidth}{!}{
		\begin{tabular}{l|ccccc}
			\toprule
			& Direct & Direct ($\alpha{=}\frac{1}{|A|}$) & Indirect & Indirect ($\alpha{=}\frac{1}{|A|}$) & Least-frequent-first\\
			\midrule
			Direct & 17 & 5 & 1 & 2 & 0\\
			Direct ($\alpha{=}\frac{1}{|A|}$) & 5 & 17 & 1 & 1 & 3\\
			Indirect & 1 & 1 & 17 & 17 & 3\\
			Indirect ($\alpha{=}\frac{1}{|A|}$) & 2 & 1 & 17 & 17 & 3\\	
			Least-frequent-first & 0 & 3 & 3 & 3 & 17\\
			\bottomrule
	\end{tabular}}
	\label{tab:tau_test_between_filters}
\end{table*}

Figure \ref{fig:average_winning_number} shows the average winning number $\overline{W}_i^x$ for different values of $x$ and for the seven different activity filtering techniques. We observe that for higher ratios of activities explained the differences between filtering techniques are smaller than for lower numbers of activities explained. Intuitively this can be explained by the fact that for lower ratios of activities explained more activities have been filtered out from the log. Therefore the effect of the filtering techniques is more clearly visible. The figure shows that, up until +-74\% of activities explained, the indirect entropy-based activity filtering technique leads to the most deterministic process models averaged over all event logs included in the experiment, where it outperforms between 4 and 4.5 other filtering techniques. Between +-75\% and +- 87.5\% the indirect entropy-based activity filtering technique with Laplace smoothing results in the highest average winning number, although the difference with the indirect entropy-based filtering technique seems negligible. Filtering out random activities from the event log outperforms none of the 6 other activities filtering techniques for the most of the graph, indicating that frequency-based filtering clearly outperforms filtering random activities.

To investigate to what degree the order in which activities are removed from the logs differs between the activity filtering techniques we calculate Kendall's tau ($\tau_b$) rank correlation for each log between the activity filtering techniques in a pairwise way. Table \ref{tab:kendall_correlation_between_filters} shows the rank correlation values found between the activity filters, averaged over the 17 event logs. The indirect activity filter with Laplace smoothing and the indirect activity filter without Laplace smoothing generate orderings over the activities of a log that are strongly correlated. Between the direct activity filter without Laplace smoothing and the direct activity filter without Laplace smoothing there is only a weak correlation. All the other activity filtering techniques are uncorrelated or very weakly correlated. Using the Kendall $\tau_b$ statistic, we apply a tau test for each pair of activity filtering techniques on each event log to test the null hypothesis that the two orderings in which activities are filtered by the two activity filtering techniques are uncorrelated, using a significance level $\alpha=0.05$.

For each pair of activity filtering techniques Table \ref{tab:tau_test_between_filters} shows the number of event logs for which the null hypothesis was rejected, i.e., the number of event logs for which the order in which activities are filtered is statistically correlated. The indirect activity filters with and without Laplace smoothing create correlated orderings of activities for all seventeen event logs. For all other pairs of activity filtering techniques the orderings in which activities are filtered are only correlated with for low numbers of event logs.

\section{Entropy-based Toggles for Process Discovery}
\label{sec:slider}

In the previous section we have shown that all four configurations of the entropy-based activity filtering technique lead to more deterministic process models compared to simply filtering out infrequent activities. However, the differences in determinism of the process models that are discovered after applying any of the four configurations are small and dependent on the event log to which they are applied. Furthermore, all four configurations of the activity filtering technique simply impose an ordering over the activities, but do not specify at which step the filtering should be stopped. Additionally, the proposed filtering technique ignores the semantics of activities: activities that are chaotic may still be relevant for the process. Leaving them out of the process model to discover will harm the usefulness of the discovered process model.

\begin{figure*}
	\centering
	\includegraphics[width=\linewidth]{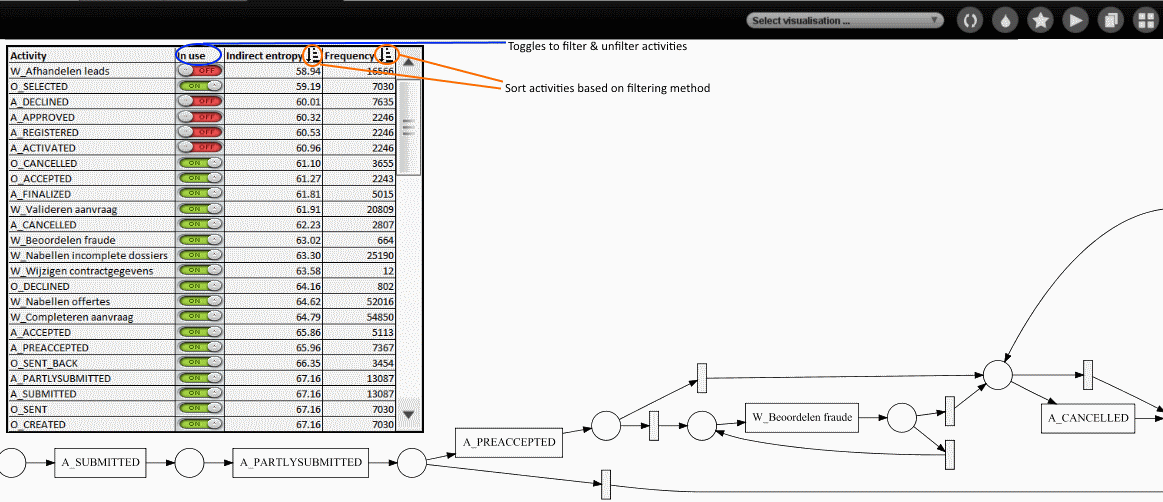}
	\caption{A mockup of the proposed way to use the activity filters in an interactive setting.}
	\label{fig:inductive_visual_miner}
\end{figure*}

To address the three issues we propose to use the filtering technique as a sorting technique over the activities in combination with toggles that interactively allow the process analyst to ``disable'' (filter out) or ``enable'' activities, and then rediscover and visualize the process model according to the new settings. This approach is similar to the Inductive Visual Miner \cite{Leemans2014}, an interactive implementation of the Inductive Miner \cite{Leemans2013} algorithm which allows the process analyst to filter the event log interactively using a slider-based approach. The Inductive visual miner contains two sliders: with one slider activities can be filtered using the least-frequent-first filter, where the user can control how many activities are filtered out by moving the slider up and down. We propose to replace this slider with a sorted list of activities and toggles, as this allows the process analyst to override the ordering of the activities that is determined by the activity filtering technique with domain knowledge. Figure \ref{fig:inductive_visual_miner} shows a mockup of the proposed way to use the activity filter. Activities are by default sorted using the chaotic activity filter, showing the entropy to indicate the assessed degree of chaoticness of each activity. Based on this information, the process analyst can choose to rely on the filtering technique and filter out the top of the list or to override this list with domain knowledge. Furthermore, other activity filtering techniques, such as the least-frequent-first filter, can be included as an additional column on which the activities of the process can be sorted. This allows the process analyst to control how many activities, and which activities, are filtered out of the process model, and thereby also empowers the user to prevent the removal of semantically important activities that should not be removed. Furthermore, this approach allows the process analyst to explore himself which of the filtering techniques leads to the most useful process model from the event log that he is analyzing.

\section{Related Work}
\label{sec:related_work}
Real life events logs often contain all sorts of data quality issues \cite{Suriadi2017}, include incorrectly logged events, events that are logged in the wrong order, and events that took place without being logged. Instances of such data quality issues are often referred to as \emph{noise}. Many event log filtering techniques have been proposed to address the problem of noise. Existing filtering techniques in the process mining field can be classified into four categories: 1) event filtering techniques, 2) process discovery techniques that have an integrated filtering mechanism build in, 3) trace filtering techniques, and 4) activity filtering techniques. We use these categories to discuss and structure related work.

\subsection{Event filtering}
Conforti et al. \cite{Conforti2017} recently proposed a technique to filter out outlier events from an event log. The technique starts by building a prefix automaton of the event log, which is minimal in terms of the number of arcs in the automaton, using an Integer Linear Programming (ILP) solver. Infrequent arcs are removed from the minimal prefix automaton, and finally, the events belonging to removed arcs are filtered out from the event log.

Lu et al. \cite{Lu2015} advocate the use of \emph{event mappings} \cite{Lu2014} to distinguish between events that are part of the mainstream behavior of a process and outlier events. Event mappings compute similar behavior and dissimilar behavior between each two executions of the process as a mapping: the similar behavior is formed by all pairs of events that are mapped
to each other, whereas events that are not mapped are dissimilar behavior.

Fani Sani et al. \cite{FaniSani2017} proposes the use of sequential pattern mining techniques to distinguish between events that are part of the mainstream behavior and outlier events.

All three of the event filtering techniques listed above aim filter out outlier events from the event log, while keeping the mainstream behavior. Event filtering techniques model the frequently occurring contexts of activities and filter out the contexts of activities that occur infrequently in the log. For example, consider an activity $B$ such that 98\% of its occurrences are in context $\langle\dots,A,B,C,\dots\rangle$, with the remaining 2\% of the events of activity $B$ are in context $\langle\dots,D,B,E,\dots\rangle$, then the $B$ events that occur between $D$ and $E$ will be filtered out by event filtering techniques. Note that our filtering technique is orthogonal to event filtering: it would consider activity $B$ to be nonchaotic and would not filter out anything. However, when a log $L$ contains a chaotic activity $X$, then event filtering techniques are not able to remove all events of this chaotic activity. One of the contexts of $X$ will by chance be more frequent than other contexts, i.e., for some activity $A$, it will hold that $\forall{B\in\mathit{Activities(L)}}:\#(\langle A,X\rangle, L)>\#(\langle B,X\rangle, L)$, even though $\langle A,X\rangle$ might only be slightly more frequent. This will result in $X$ events after a $B$ being removed, while the $X$ events after an $A$ remain in the log. Applying a process discovery technique to this filtered log will then result in a process model where activity $X$ is misleadingly positioned after activity $A$, while in fact $X$ can happen anywhere in the process. The activity filtering technique presented in this paper will instead detect that activity $X$ is chaotic, and completely remove it from the event log, preventing the misleading effect of event filtering.

\subsection{Process Discovery Techniques with Integrated Filtering}
Several process discovery algorithms offer integrated filtering mechanisms as part of the approach. The Inductive Miner (IM) \cite{Leemans2013b} is a process discovery algorithm which first discovers a \emph{directly-follows graph} from the event logs, where activities are connected that directly follow each other in the log, from which in a second step a process model is discovered. The directly-follows relations are affected by the presence of a chaotic activity $X$: sequence $\langle\dots,A,X,C,\dots\rangle$ leads to false directly-follows relations between $A$ and $X$ and between $X$ and $C$, while the directly-follows relation between $A$ and $C$ is obfuscated by $X$. The Inductive Miner infrequent (IMf) \cite{Leemans2013} is an extension of the IM where infrequent directly-follows relations are filtered out from the set of directly-follows relations that are used to generate to process models. The filtering mechanism of IMf can help to filter out the directly-follows relations between $A$ and $X$ and between $X$ and $C$, but it does not help to recover the obfuscated directly-follows relation between $A$ and $C$. Instead, the activity filtering technique presented in this paper filters out the chaotic activity $X$, leading to sequence $\langle\dots,A,X,C,\dots\rangle$ being transformed into $\langle\dots,A,C,\dots\rangle$, thereby recovering the directly follows relation between $A$ and $C$.

The Heuristics Miner \cite{Weijters2011} and the Fodina algorithm \cite{Fodina2017}, in addition to the directly-follows relation, defines an \emph{eventually-follows relation} between activities and allows the process analyst to filter out infrequent directly-follows and eventually follows relations. Two activities $A$ and $B$ are in an eventually-follows relation when $A$ is eventually followed by $B$, before the next appearance of $A$ or $B$. The eventually-follows relation, unlike the directly-follows relation, is not impacted by the presence of chaotic activities.
The Heuristic Miner \cite{Weijters2011} and Fodina \cite{Fodina2017} both include filtering methods for the directly-follows and eventually-follows relations that are similar in nature to the filtering mechanism that is used in the Inductive Miner infrequent \cite{Leemans2013}. However, the use of sequential orderings and parallel constructs in the mining approaches of the Heuristic Miner \cite{Weerdt2011} and Fodina \cite{Fodina2017} is based on the directly-follows relations only, with the eventually follows relations being used for the mining of long-term dependencies. Furthermore, in contrast to the Inductive Miner, the process models discovered with the Heuristic Miner \cite{Weijters2011} or Fodina \cite{Fodina2017} can be unsound, i.e., the can contain deadlocks.

The ILP-miner \cite{Werf2009} is a process discovery algorithm where a set of behavioral constraints over activities is discovered for each prefix (called the \emph{prefix-closure}) of the event log, based on which a process model is discovered that satisfies these constraints using \emph{Integer Linear Programming} (ILP). Van Zelst et al. \cite{Zelst2015} proposed a filtering technique for the ILP-miner where the prefix closure of the event log is filtered prior to solving the ILP problem by removing infrequently observed prefixes. It is easy to see that a chaotic activity $X$ affect the prefix-closure that is discovered from the event log: given log consisting of two traces $\langle A,X,C\rangle$ and $\langle X,A,C\rangle$, activity $X$ causes the prefixes closures of the two traces to have no overlap in states, while without activity $X$ the two traces are identical. This makes the filtering method of the prefix-closure proposed by Van Zelst et al. \cite{Zelst2015} less effective, as frequent prefixes randomly get distributed over several infrequent prefixes when chaotic activities are present. Instead, the chaotic activity filtering technique presented in this paper would remove chaotic activity $X$, leading to traces $\langle A,X,C\rangle$ and $\langle X,A,C\rangle$ becoming identical after filtering, therefore leading to a simpler process model while still describing the behavior of the event log accurately.

The Fuzzy Miner \cite{Gunther2007} is a process discovery algorithm that aims at mining models from flexible processes, and it discovers a process model without formal semantics. The Fuzzy Miner discovers this graph by extracting the eventually follows relation from the event log, which is not affected by chaotic activities. Similar to the Heuristics Miner \cite{Weijters2011} and Fodina \cite{Fodina2017} the Fuzzy Miner allows to filter out infrequent eventually-follows relations between activities. In practice, the lack of formal semantics of the Fuzzy Miner models hinders the usability of the models, as the models are not precise on what behavior is allowed in the process under analysis.

\subsection{Trace filtering}
Ghionna et al. \cite{Ghionna2008} proposed a technique to identify outlier traces from the event log that consists of two steps: 1) mining frequent patterns from the event log, and 2) applying MCL clustering \cite{Dongen2008} on the traces, where the similarity measure for traces is defined on the number of patterns that jointly characterize the execution of the traces. Traces that are not assigned to a cluster by the MCL clustering algorithm are considered to be outlier traces and are filtered from the event log. It is easy to see that trace filtering techniques address a fundamentally different problem than chaotic activity filtering: in the event log shown in Figure \ref{sfig:local_model_example_noisy} there are only two traces that do not contain an instance of chaotic activity $X$, therefore, even if a trace filtering technique would be able to perfectly filter out traces that contain a chaotic event, the number of remaining traces will become too small to mine a fitting and precise process model when the chaotic activity is frequent.

\subsection{Activity filtering}
The modus operandi for filtering activities is to simply filter out infrequent activities from the event log. The plugin \emph{'Filter Log using Simple Heuristics'} in the ProM process mining toolkit \cite{Dongen2005} offers tool support for this type of filtering. The Inductive Visual Miner \cite{Leemans2014} is an interactive process discovery tool that implements the Inductive Miner \cite{Leemans2013} process discovery algorithm in an interactive way: the process analyst can filter the event log using sliders and is then shown the process model that is discovered from this filtered log. One of the available sliders in the Inductive Visual Miner offers the same frequency-based activity filtering functionality. The working assumption behind filtering out infrequent activities is that when there are just a few occurrences of an activity, there is probably not enough evidence to establish their relation to other activities to model their behavior. However, as we have shown in this paper, for frequent but chaotic activities, while they are frequent enough to establish their relation to other activities, complicate the process discovery task by lowering directly-follows counts between other activities in the event log. The activity filtering technique presented in this paper is able to filter out chaotic activities, thereby reconstructing the directly-follows relations between the non-chaotic activities of the event log, at the expense of losing the chaotic activities.

\section{Conclusion \& Future Work}
\label{sec:conclusion}
In this paper, we have shown the possible detrimental effect of the presence of chaotic activities in event logs on the quality of process models produced by process discovery techniques. We have shown through synthetic experiments that frequency-based techniques for filtering activities from event logs, which is currently the \emph{modus operandi} for activity filtering in the process mining field, do not necessarily handle chaotic activities well. As shown, chaotic activities can be frequent or infrequent. We have proposed four novel techniques for filtering chaotic from event logs, which find their roots in information theory and Bayesian statistics. Through experiments on seventeen real-life datasets, we have shown that all four proposed activity filtering techniques outperform frequency-based filtering on real data. The indirect entropy-based activity filter has been found to be the best performing activity filter overall averaged over all datasets used in the experiments; however, the performance of the four proposed activity filtering techniques is highly dependent on the characteristics of the event log.

Because the performance of the filtering techniques was found to be log-dependent, we propose the use the activity filtering techniques in a slider-based approach where the user can filter activities interactively and directly see the process model discovered from the filtered event log. Ultimately, only the user can decide which activities to include. In future work, we aim to construct a hybrid activity filtering technique that combines the four techniques proposed in this paper by using supervised learning techniques from the data mining field to predict the effect of removing a particular activity.


%


\bibliographystyle{spbasic}      
\bibliography{bibliography}



\end{document}